\documentclass[journal,twoside]{IEEEtran}

\usepackage{amsmath}
\usepackage{amssymb}
\usepackage{latexsym}
\usepackage{graphicx}
\usepackage{epsfig}
\usepackage{psfrag}
\usepackage[active]{srcltx} 
\usepackage{color}
\usepackage{setspace}
\usepackage[normal,footnotesize]{caption}
\usepackage{balance}
\usepackage{textcomp}
\usepackage{url}
\usepackage{subfig}

\newtheorem{theorem}{\bf{Theorem}}
\newtheorem{corollary}{\bf{Corollary}}

\newtheorem{remark}{\it{Remark}}

\newcommand{\insertonefig}[4]
{
\begin{figure}[htbp]
\begin{center}
\includegraphics[width=#1,clip,keepaspectratio]{#2}
\end{center}
\caption{#4}
{#3}
\end{figure}
}

\newcommand{\inserttwofigsV}[5]
{
\begin{figure}[h]
\begin{center}$
\begin{array}{c}
\includegraphics[width=#1,clip,keepaspectratio]{#2} \\
\text{(a)} \\
\includegraphics[width=#1,clip,keepaspectratio]{#3} \\
\text{(b)} \\
\end{array}$
\end{center}
\caption{#5}
{#4}
\end{figure}
}

\newcommand{\addphoto}[1]{\includegraphics[width=1in,height=1.25in,clip,keepaspectratio]{#1}}

\newcommand{\noi} {\noindent}

\newcommand{\bA}{\mathbf{A}}
\newcommand{\bbA}{\boldmath{\mbox{$A$}}}

\newcommand{\bB}{\mathbf{B}}
\newcommand{\bbB}{\boldmath{\mbox{$B$}}}

\newcommand{\bC}{\mathbf{C}}
\newcommand{\bbC}{\boldmath{\mbox{$C$}}}

\newcommand{\bD}{\mathbf{D}}
\newcommand{\bbD}{\boldmath{\mbox{$D$}}}

\newcommand{\bh}{\mathbf{h}}
\newcommand{\bH}{\mathbf{H}}
\newcommand{\bbH}{\boldmath{\mbox{$H$}}}
\newcommand{\bi}{\mathbf{i}}
\newcommand{\bI}{\mathbf{I}}

\newcommand{\bT}{\mathbf{T}}

\newcommand{\bv}{\mathbf{v}}

\newcommand{\bx}{\mathbf{x}}

\newcommand{\bzero}{\mathbf{0}}

\newcommand{\bXi}{\mbox{\boldmath $\Xi$}}

\begin{document}


\title{Discrete-Time Block Models\\for Transmission Line Channels:\\Static and Doubly Selective Cases}

\author{

        Stefano~Galli,~\IEEEmembership{Senior~Member,~IEEE,}
        Anna~Scaglione,~\IEEEmembership{Fellow,~IEEE}
\thanks{S. Galli (sgalli@ieee.org) is is with ASSIA, Inc., 333 Twin Dolphin Drive, Redwood City, CA
94065. At the time of writing this paper, Dr. Galli was with Panasonic Corporation.}

\thanks{A. Scaglione (ascaglione@ucdavis.edu) is with the ECE Dpt., University of California,
Davis, CA 95616, USA.}

\thanks{The paper is under review with the IEEE Trans. on Commun.}

} 

\markboth{Submitted to the IEEE Transactions on Communications, June~2009. Revised Nov. 2009}%
{S. Galli \MakeLowercase{\textit{et al.}}: Discrete-Time Block Models for Transmission \ldots}


\maketitle

\begin{singlespace}
\begin{abstract}

\noi Most methodologies for modeling Transmission Line (TL) based channels define the input-output relationship in the frequency domain (FD) and handle the TL resorting to a two-port network (2PN) formalism. These techniques have not yet been formally mapped into a discrete-time (DT) block model, which is useful to simulate and estimate the channel response as well as to design optimal precoding strategies. TL methods also fall short when they are applied to Time Varying (TV) systems, such as the power line channel. The objective of this paper is to establish if and how one can introduce a DT block model for the Power Line Channel. We prove that it is possible to use Lifting and Trailing Zeros (L\&TZ) techniques to derive a DT block model that maps the TL-based input-output description  directly in the time domain (TD) block channel model. More specifically, we find an interesting relationship between the elements of an ABCD matrix, defined in the FD, and filtering kernels that allow an elegant representation of the channel in the TD. The same formalism is valid for both the Linear Time Invariant (LTI) and the Linear TV (LTV) cases, and bridges communications and signal processing methodologies with circuits and systems analysis tools.

\end{abstract}
\end{singlespace}


\thispagestyle{empty}

\section{Introduction}

\IEEEPARstart{T}{oday} many Home Networking (HN) technologies are available to the consumer. The general consensus is that HN is moving beyond simple data sharing among PCs and their peripherals and that multimedia applications will be at the heart of HN products.  Although wireless Local Area Networks (LAN) based on the IEEE 802.11 standard are the most popular HN solutions,  they often suffer from poor RF propagation and from mutual interference \cite{LinLatNew2003}.  Hence, ``wired'' in-home (IH) connections are gaining new momentum, and high-speed HN technologies over IH power lines (PLs), phone lines (PHs), and coaxial cables (CX) are being intensely investigated. Among them, Power Line Communications (PLCs) represents the most accessible as well as the most challenging medium \cite{Big2003}, \cite{Book:Dostert2001}. Initially, the interest for PLCs was targeted to using the existing outdoor PL infrastructure for delivering broadband Internet access \cite{GalScaDos2003}. More recently, PLC technology is emerging as an excellent candidate for many other applications, e.g. for in-home, in-office and in-vehicle LANs, for smart grid applications, and in building control and automation networks \cite{PavHanYaz2003}, \cite{LatYon2003}, \cite{BigGalLee2006}. These technologies are also being standardized in IEEE and ITU-T. PLCs are the scope of the recently confirmed IEEE P1901 baseline document  \cite{GalLog2008}, whereas ITU-T has recently consented to Recommendation G.9960 (also known as G.hn) for a unified IH networking technology operating over of all types of IH wiring \cite{OksGal2009}.

TL-based channels like PLs, PHs and CXs have been traditionally modeled in the FD representing section of cables and various common discontinuities like bridged taps, transformers, series impedances, splitters, etc., via Two-Port Networks (2PN) and their corresponding ABCD or Transmission Matrices (TMs) \cite{Wer1991}, \cite{Book:StaCioSil1999}, \cite{GalBanPII2005}, \cite{GalBan2006}. While the modeling of CX and PH is today fairly well established, the modeling of PL is still under intense debate.  The recent results published in \cite{GalBanPII2005}, \cite{GalBan2006}, confirm that the ABCD-based formalism is adequate to describe signal propagation over PLs and that this conclusion holds also when grounding is present. However, a peculiar aspect of the PL channel that differentiates it from the other two IH TL-based channels is that the PL channel is a Linear and Periodically Time Varying (LPTV) channel as demonstrated by Ca{\~n}ete et al. in \cite{CanCorDie2006} (see also Sect. \ref{sec.model-TVchannel-LTV} for more details).

Time-varying DT models for the TL-based channel are not available yet, since modeling has been traditionally carried out in the Laplace or Fourier Domain. These domains are inappropriate for time-varying systems. Some initial work on the modeling of LPTV nature of the PL channel can be found in \cite{CanDieCor2002, Cav2004, CanCorDie2006, BarmMusoTucc09} and references therein. These previous  attempts at modeling do not provide a direct mapping between the modulated symbols and the output data, which typically are used to estimate as well as optimize the precoding method. Motivated by the fact that key advances in broadband wireless and DSL technologies were fostered by utilizing block transmission models and precoding strategies, our work contribution is twofold\footnote{~Some initial results were presented at the 2008 IEEE ISPLC conference \cite{SunScaGal2008}; however, only the simpler case of Trailing Zeros was addressed there and the Zadeh decomposition was not given.}: 1) we prove that block models similar to those used in wireless and wireline DSL channels can be used in the PLC conext as well; 2) we provide a DT model capable of handling the LTI, LTV and LPTV cases under the same formalism. The block transmission model we propose is immediately useful for the analysis of present designs, since the IH networking standards use schemes based on OFDM or Wavelet-OFDM \cite{GalLog2008}, \cite{OksGal2009}, or single carrier schemes based on block DFE \cite{STD:HomePNA-2007}.

The novelty of the method proposed here is that it addresses cases not contemplated by previous work on the modeling of the TV nature of the PL channel. For example, \cite{CanCorDie2006} allows the modeling of TV cases where the time variability is such that it can be modeled as a succession of LTI states thus allowing the use of a Fourier basis, whereas the goal of \cite{BarmMusoTucc09} is limited to the estimation of upper and lower bounds on the channel impulse response variations due to TV loads. Our work aims at mapping the modulated input signal to the output, modeling the variability of the channel using integral equations and then lifting and not a Fourier basis. As we will show here, this approach naturally leads to the design of precoders and bit loading methods, as well as facilitating receiver analysis and optimization.

The paper is organized as follows.
The transition from a FD-based to a TD-based model for TL-based channels is addressed in Sect. \ref{sec.model-TVchannel-LTV} for the LTV and LPTV cases.
The lifted block representation of the DT model and the final input-output relationship of a TL-based channel is derived in Sect. \ref{sec.LiftedForm}; in this section, we also verify the finite support of the employed LTV/LPTV/LTI filtering kernels and we propose an MMSE kernel estimator.
In Sect. \ref{sec.ChainRule} we show under what conditions the useful Chain Rule valid in the FD holds also in DT lifted form.
Finally, we give conclusive remarks in Sect. \ref{sec.Conclusions}.

\subsection{Notation and Acronyms}
In the rest of the paper we use the acronyms listed in Table \ref{tab.acronyms} and use the following mathematical conventions: boldface upper (lower) case letters denote matrices (vectors); $P\times P$ channel matrices is denoted by the symbol $\bH$ and its tall counterpart obtained by removing the last $L$ columns is denoted by $\bbH$; similarly, $P\times 1$ vectors $\bv[i]$ with trailing zeros are denoted as $\bv[i]=(\underline{\bv}_s^T[i],0,\ldots,0)^T$; element $(k,n)$ of matrix $\bH$ is denoted as $\{ {\bH}\}_{k,n}$; inverse and pseudoinverse matrices are denoted as $^{-1}$ and $^\dagger$, respectively.

\begin{table*}[htbp]
  \centering
  \caption{List of acronyms used in the paper}\label{tab.acronyms}
\begin{tabular}{llcll}
  \hline\hline
  Acronym & Meaning & & Acronym & Meaning \\ [0.5ex]
  \hline
2PN	&	Two	Port Network	         &	&  L\&TZ   & Lifting and Trailing Zero\\			
AC	&	Alternate Current            &  &  LPTV	   & Linear	and	Periodically Time Variant\\		
CX	&	Coaxial	Cable                &	&  LTI     & Linear	Time Invariant\\	
DT	&	Discrete Time                &	&  LTV     & Linear	Time Variant	\\				
FD	&	Frequency Domain             &	&  OFDM    & Orthogonal	Frequency Division Multiplexing	\\			
FT	&	Fourier	Transform            &	&  PH      & Phone Line	\\		
FIR	&	Finite Impulse Response      &	&  PL      & Power Line	\\				
HN	&	Home Networking              &	&  PLC     & Power Line Communication\\				
IBI	&	Inter Block Interference     &	&  RMS-DS  & Root-Mean-Square Delay Spread\\				
ICI	&	Inter Carrier Interference   &	&  TD      & Time Domain \\				
IFT	&	Inverse	Fourier	Transform	 &	&  TL      & Transmission Line \\				
IH	&	In-Home	                     &	&  TM      & Transmission Matrix \\				
IIR	&	Infinite Impulse Response    &	&  TV      & Time Varying \\	
LAN & Local Area Network             &  &  TZ      & Trailing Zero \\ [1ex]			
  \hline
\end{tabular}
\end{table*}

\section{From Frequency Domain to Time Domain: the LTV and LPTV Cases} \label{sec.model-TVchannel-LTV}

A general result of TL theory is that every uniform TL can be modeled as a Two-Port Network (2PN), thus allowing us to replace a distributed parameter circuit with a single lumped network. A 2PN has an associated 2-by-2 TM whose four elements ($A$, $B$, $C$, and $D$) are complex functions of the frequency. The steady-state relationship between current and voltage (in the FD) at the two ports of a 2PN is tied by $\bT_f$, the forward TM (see Fig. \ref{fig.TPN00}):
\begin{equation}\label{eq.IO}
    \left[
      \begin{array}{c}
                V_{in} \\
                I_{in} \\
      \end{array}
    \right]
    =
    \bT_f
    \left[
       \begin{array}{c}
        V_{out}  \\
        I_{out}  \\
       \end{array}
    \right]
    =
    \left[
       \begin{array}{cc}
        A  & B \\
        C  & D  \\
       \end{array}
    \right]
    \left[
       \begin{array}{c}
        V_{out}  \\
        I_{out}  \\
       \end{array}
    \right].
\end{equation}
In general, a TL-based link is made of several sections; each section consists of segments of different cables of various lengths spliced together. Series/shunt impedances, bridged taps, and other discontinuities may be present along the line. The TM of the overall link is simply obtained by multiplying the respective TMs of each portion of the network (Chain Rule). The TM formalism was first used in digital communications applications by J.J. Werner for PH modeling \cite{Wer1991}. Later it was adopted by Galli and Banwell for PL channels with and without grounding \cite{GalBan2006, BanGal2001}, and by Chen for the CX case \cite{Book:Chen2003}.

\insertonefig{8.8cm}{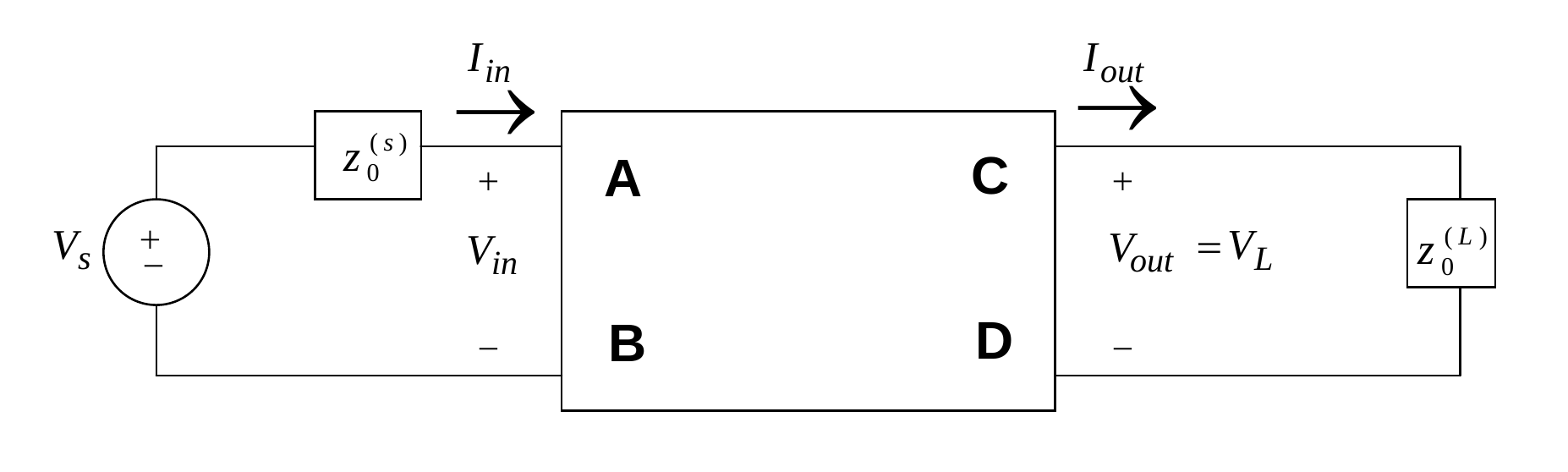}{\label{fig.TPN00}}
{Schematic representation of a two-port network (2PN).}

It is useful to express output quantities as a function of input quantities inverting \eqref{eq.IO}:
\begin{equation}\label{eq.IOinv}
    \left[
      \begin{array}{c}
                V_{out} \\
                I_{out} \\
      \end{array}
    \right]
    =
    \bT_b
    \left[
       \begin{array}{c}
        V_{in}  \\
        I_{in}  \\
       \end{array}
    \right]
\end{equation}
where the backward TM is $\bT_b=(\bT_f)^{-1}$.

The transfer function of the 2PN, i.e. the ratio of the voltage on the load $z^{(L)}_0$ to the source voltage $z^{(s)}_0$, can be easily expressed in terms of ABCD parameters:
\begin{eqnarray} \label{eq.Hf}
    H(f_0) &=& \frac{V_L(f_0)}{V_S(f_0)} \\
           &=& \frac{z^{(L)}_0}{Az^{(L)}_0 +B+Cz^{(L)}_0 z^{(s)}_0+Dz^{(s)}_0}
\end{eqnarray}
For some notable cases, the elements of the forward TM can be written in closed form:
\begin{equation}\label{eq.ABCDdef}
\begin{cases}
\text{Cable of length $l$:} & A = D = \cosh({\gamma(f) l}); \\
\text{}                     & B = Z_{o}(f) \sinh(\gamma(f) l); \\
\text{}                     & C = \sinh(\gamma(f) l)/Z_{o}(f) \\
\text{Shunt impedance $Z$:} & A = D = 1 ; ~~B = 0 ;~~~C = 1/Z \\
\text{Series impedance $Z$:} & A = D = 1 ; ~~B = Z ;~~~C = 0 \\
\end{cases}
\end{equation}
where $\gamma(f)$ and $Z_{o}(f)$ are respectively the propagation constant and the characteristic impedance of the cable. The behavior of $\gamma(f)$ and $Z_{o}(f)$ for two typical IH wires is shown in Fig. \ref{fig.GammaZo}; $\gamma(f)$ grows very quickly with frequency whereas $Z_{o}(f)$ can be considered approximately constant in the High Frequency (HF) band and above. Furthermore, $\det(\bT_f)=\det(\bT_b)=1$, i.e. the 2PN models of TLs and of shunt/series  impedances are reciprocal 2PNs. In this case, \eqref{eq.IO} becomes:
\begin{equation}\label{eq.IO5}
\begin{cases}
V_{out}(f)  = D(f) V_{in}(f)-B(f)I_{in}(f)  \\
I_{out}(f)  = -C(f) V_{in}(f)+A(f)I_{in}(f)
\end{cases}
\end{equation}
Starting from the above FD representation of a TL, we will now introduce our DT models starting from the general LTV case and then particularizing it to the LPTV case. Then in Sect. \ref{sec.LiftedForm} we will introduce the proposed block form representation (lifting) and show how the LTV, LPTV and LTI cases can be all handled under the same formalism.

\insertonefig{8.8cm}{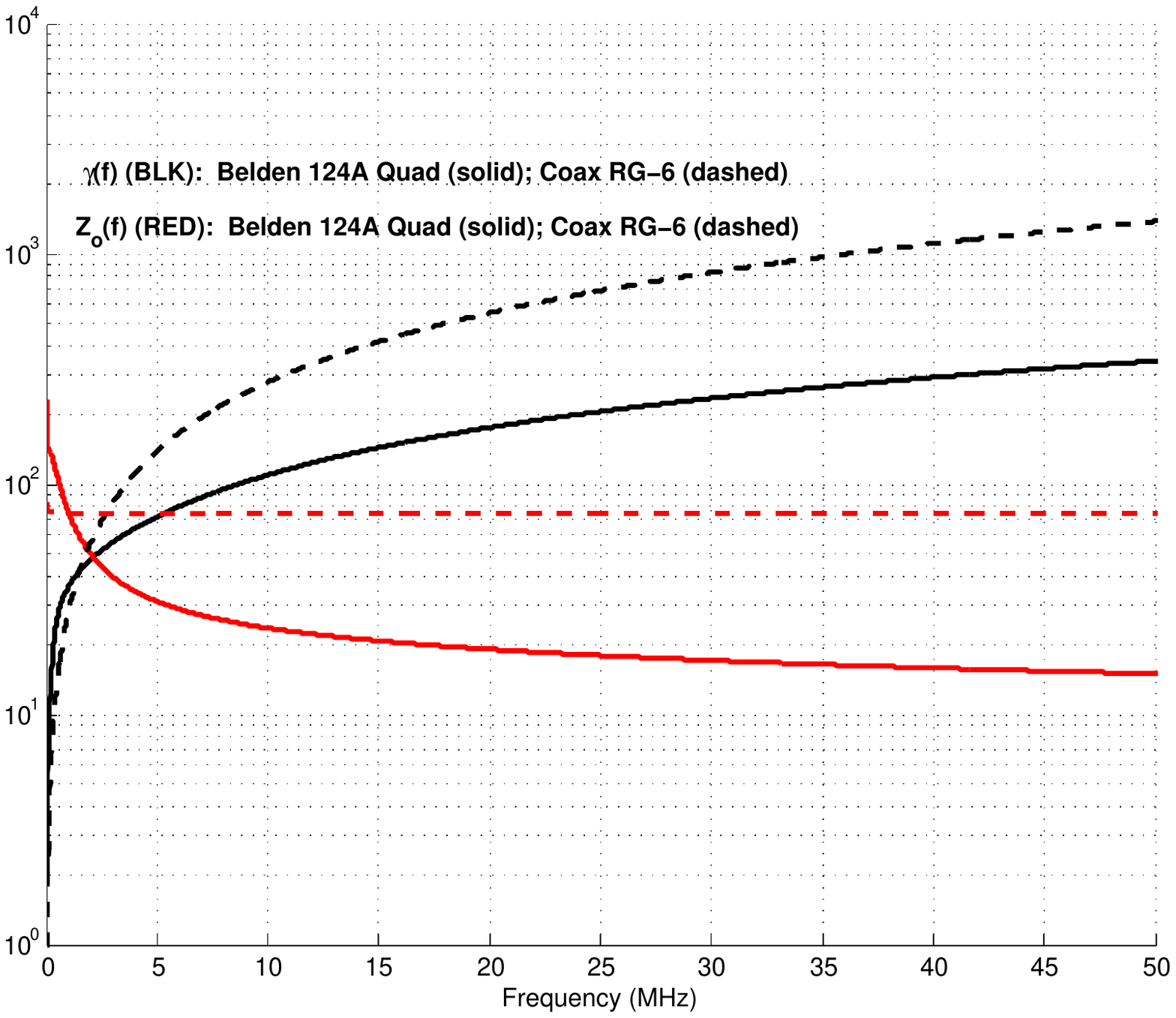}{\label{fig.GammaZo}}
{Plot of the absolute value of $\gamma (f)$ and $Z_{o}(f)$ versus frequency for two typical IH wires.}

\subsection{The LTV Case} \label{sec.LTVcase}

In the LTV case \eqref{eq.IO5} does not hold because TV systems exhibit integral input-output equations in the frequency domain as well as in the time-domain. In fact, since the Fourier basis is no longer the basis of eigenfunctions in the case of LTV systems, we argue that there is no particular advantage in analyzing the system in frequency versus time. Furthermore, since communication signals are band-limited, a discrete time description is appropriate and sufficient to describe the input output relationships. Our objective is to lay down DT models that are flexible enough to handle the LTV, LPTV, and LTI cases.

Since we are in the linear regime, in general we can write:
\begin{align}
v_{out}(t) &= {\cal L}_v[ v_{in}(t), i_{in}(t)] \nonumber \\
i_{out}(t) &=  {\cal L}_i[ v_{in}(t), i_{in}(t)]
\end{align}
Hence, it is still possible to view the electrical quantities on Port 2 as filtered versions of the quantities on Port 1, where now the filters are represented by TV kernels, analogous to the time-varying impulse response introduced by Bello  \cite{Bello1964}, as follows:
\begin{IEEEeqnarray}{rCl}
v_{out}(t) &=& \int_{-\infty}^{\infty} v_{in}(\tau) d(t,t-\tau) d\tau \\
    && -\: \int_{-\infty}^{\infty} i_{in}(\tau) b(t,t-\tau) d\tau \nonumber  \\
i_{out}(t) &=& -\int_{-\infty}^{\infty} v_{in}(\tau) c(t,t-\tau) d\tau \\
    && +\: \int_{-\infty}^{\infty} i_{in}(\tau) a(t,t-\tau) d\tau \label{eq.tvconv11b} \nonumber,
\end{IEEEeqnarray}
where $d(t,t-\tau)={\cal L}_v [\delta(t-\tau),0]$, $-b(t,t-\tau)={\cal L}_v [0,\delta(t-\tau)]$, $-c(t,t-\tau)={\cal L}_i [\delta(t-\tau),0]$, and $a(t,t-\tau)={\cal L}_i [0,\delta(t-\tau)]$.
Our work assumes that an expression for the above TV kernels is available, but we point out that it is not trivial to obtain TV kernels in complete generality. Finding an appropriate mathematical representation is, however, a first essential ingredient to obtain the kernels either experimentally or through circuit analysis.

There are differences between time variations caused by non uniform TLs and TV loads with uniform TLs. In \cite{Weinstein1965}, \cite{EllGun1966}, the authors studied the case of non uniform TLs where the time variations are due to changes of the primary parameters of the TL. This case is seldom of interest since the most common TL based communications channels owe their time variability to loads only. The case of greatest interest in the context of data communications is when the time-variabilty of the loads are TV and the time-variability is periodic (LPTV).

\subsection{Zadeh's Expansion Approach for the LPTV Case and Its Extensions} \label{sec.LPTVcase}

The IH PL channel measurements in \cite{CanCorDie2006} give insight on how many significant harmonic components are present in the PL channel. \cite{CanCorDie2006} reported a median Dopppler spread of 100 Hz, a $90\%$-percentile of 400 Hz, and Doppler components up to $B_D^{(max)}=1,750$ Hz; thus, the coherence time of the channel is around $1/B_D^{(max)} \approx 600 \mu s$. Interestingly, all observed Doppler components were quantized at multiples of the fundamental frequency of the mains AC cycle, i.e. $f_0=1/T_0=50Hz$.  This confirms that the PL channel can be modeled as an LPTV channel. The LPTV nature of the PL channel has already been experimentally confirmed in \cite{CanDieCor2002, CanCorDie2006} and this paper is not concerned with confirming the validity of the LPTV model. Rather, our objective is that of finding a convenient representation and DT model for the overall channel response for the design and analysis of communication techniques over the PL channels.

An LPTV system $y(t)={\cal L}[x(t)]$ with period $T_0$ is such that its time-varying impulse response ${\cal L}[\delta(t-\tau)]=h(t,t-\tau)$ is periodic with respect to time, i.e. $h(t,\xi)=h(t+kT_0,\xi), \forall k\in {\mathbb Z}$. In \cite{Zadeh1950}, Zadeh introduced the following expansion:
\begin{equation}\label{eq.ctmodel-chanFS}
h(t,\tau)=\sum_{m=-\infty}^{+\infty} h_m(\tau)e^{j 2 \pi m f_0 t}
\end{equation}
where the so-called $harmonic$ impulse responses are defined as below:
\begin{equation}\label{eq.ctmodel-chanFS2}
h_{m}(\tau)=\frac{1}{T_0}\int_{0}^{T_0} h(t,\tau)e^{- j 2 \pi m f_0 t}dt
\end{equation}
In Zadeh's expansion an LPTV channel is equivalently represented as a bank of LTI channels whose outputs are modulated by Fourier harmonics with frequencies that are integer multiples of the fundamental frequency $f_0$ (see Fig. \ref{fig.LPTV}):
\begin{eqnarray}\label{eq.iolptv}
y(t) &=& \sum_{m=-\infty}^{+\infty} e^{j 2 \pi m f_0 t} \int_{-\infty}^{\infty} h_m(\tau) x(t-\tau) d\tau \\
     &=& \sum_{m=-\infty}^{+\infty} e^{j 2 \pi m f_0 t} \{ h_m(\tau) \ast x(t-\tau) \}
\end{eqnarray}

The expansion given in \eqref{eq.ctmodel-chanFS2} uses the orthonormal Fourier basis $\{1/\sqrt{T_0} e^{j 2 \pi m f_0 t/T}rect_{T}(t)\}_{m=-\infty}^{\infty}$. For such an orthonormal basis, if only $M$ terms are used for estimating the channel response, then the channel estimate $\hat h(t,\tau)=\sum_{m=-M}^M h_m(\tau)e^{j 2 \pi m f_0 t}$ incurs a truncation error $E^{(M)}$ that is proportional to the energy of the truncated terms:
\begin{equation}\label{eq.TruncationFourier}
    E^{(M)} = \sum_{|m|>M} \int |h_m(\tau)|^2 d\tau.
\end{equation}
\insertonefig{8.8cm}{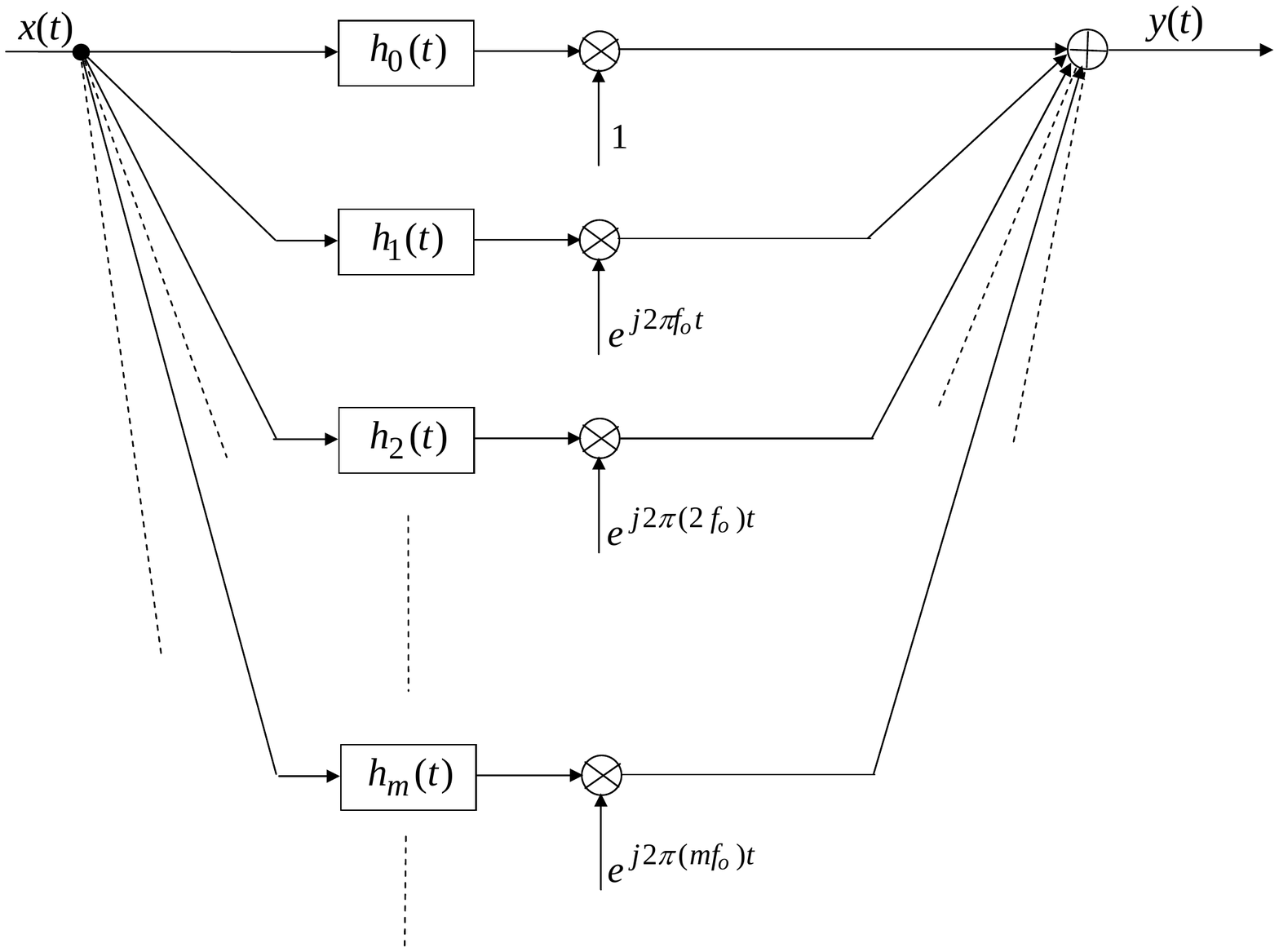}{\label{fig.LPTV}}
{Series expansion of an LPTV channel.}
\begin{remark}
It is desirable to select a basis for which the summation in \eqref{eq.TruncationFourier} grows slowly, i.e. a sparse basis. In the case of relatively abrupt discontinuities, the truncated terms of the Fourier basis in \eqref{eq.ctmodel-chanFS2} die out slowly. In these cases, one may replace Zadeh's model, with a more general basis expansion model for the TV response in the period $h_{T_0}(t,\tau)=h(t,\tau)rect_{T_0}(t)$, and obtaining back the $h(t,\tau)=\sum_{p=-\infty}^{\infty}h_{T_0}(t-pT_0,\tau)$. The expansion of $h_{T_0}(t,\tau)$ is analogous to \eqref{eq.ctmodel-chanFS} and obtained replacing with a Wavelet basis $\{p_{m}(t)\}_{m=-\infty}^{\infty}$ the orthonormal Fourier basis $\{1/\sqrt{T_0} e^{j 2 \pi m f_0 t/T}rect_{T}(t)\}_{m=-\infty}^{\infty}$. The advantage is that, if appropriately chosen, the basis $\{p_{m}(t)\}_{m=-\infty}^{\infty}$ can concentrate the energy of the channel response in fewer coefficients $h_m(\tau)$ \cite{Mallat:98}. For an orthonormal expansion, the $h_m(\tau)$ would be simply replaced by the following responses/coefficients $h_{m}(\tau)=\int_{0}^{T_0} h(t,\tau)p^*_m(t)dt$. Seeking alternatives to the Zadeh's expansion is useful to reduce the truncation error that is inevitable in estimating or simulating the channel through these models.
\end{remark}
\begin{remark}
Note also that the presence of Dopplers in Zadeh's model for an LPTV system suggests that multicarrier transmission schemes will be inevitably affected by inter-carrier interference (ICI), regardless of the duration of the guard interval. ICI will cause greater performance degradation in conventional or windowed OFDM than in Wavelet-OFDM \cite{GalKogKod2008} or OFDM/OQAM \cite{SkrSioJav07} since Gaussian shaped filters in filterbanks are more effective at rejecting ICI.
\end{remark}

Several methods have been reported to model LPTV 2PNs (see \cite{BarSab1972, Kurth1977, TohKok1996}). The first two generalize the ABCD TM formalism, considering LPTV responses, which have, however, no memory. The third reference \cite{TohKok1996} derives the general response to an exponential waveform of a network that has periodically varying elements with the intent of extending the traditional spectral analysis methods to the LPTV case. More recently \cite{VanGieSan2002} proposed to resort to a signals' sub-band decomposition, in the form $x(t)=\sum_{k=-\infty}^{+\infty} x_k(t)e^{j2\pi \frac k T t}$, along with Zadeh's filterbank expansion of the network response \cite{Zadeh1950}. Overall the authors express the linear relationship between the input and output subband frequency components using an infinite size Toeplitz matrix (called Harmonic Transfer Matrix) whose elements are $H_{n-m}(f+m f_0)$, where $H_m(f)$ is the FT of Zadeh's $h_m(t)$. Furthermore, \cite{VanGieSan2002} also presents an algorithm for the generation of symbolic expressions for the harmonic transfer functions of an LPTV systems starting from a system model in the form of a block diagram. What is inconvenient about this expansion, is that it requires the expansion of $x(t)$ as $\sum_{k=-\infty}^{+\infty} x_k(t)e^{j2\pi \frac k T t}$ and the $x_k(t)$ have no particular relationship to the encoding and modulation of the signals.

In general, a key advantage and difference of our representation is that we combine different bases expansions for input and channel. In fact, the basis chosen for the expansion of the LPTV channel does not in general have to be used as the basis of representation for the input signal as well, which is naturally represented through its samples, given that it is a band-limited communication signal. Furthermore, due to receiver filtering, the received data are also better represented through their samples, and many of the off bandwidth terms of the Zadeh's expansion are eliminated from the signal through filtering. Hence, the main contribution of this paper is to provide a new representation for the channel, that is consistent with the LPTV model but that, compared to the previous literature on the subject, operates directly in the sampled time domain.  Our representation is centered on the idea that the memory of the LPTV PLC channels is approximately finite and that we can conclude that the harmonic responses are also of finite memory. Note that is is a sufficient but not necessary condition to have the overall LPTV PLC channel have finite memory.

We verify this hypothesis in Section \ref{sec.LTIcase} relying on the fact that time-variations happen at a slow time scale compared to the ordinary symbol rates of PLC modems. That is reasonable, since the LPTV behavior is due to the fact that the electrical devices plugged in outlets (loads) contain non-linear elements such as diodes and transistors that, relative to the small and rapidly changing communication signals, appear as a resistance biased by the AC mains voltage. AC adaptors in particular, are highly likely to exhibit an LPTV input impedance. In fact, the periodically changing AC signal swings the devices over different regions of their non-linear I/V curve and this induces a periodically TV change of their resistance. The overall impedance appears as a shunt impedance across the ``hot'' and ``return'' wires and, since its time variability is due to the periodic AC mains waveform, it is naturally periodic. Furthermore, electrical devices are noise generators and, in view of Nyquist theorem, noise also appears to be cyclostationary.

Although in all cases of practical interest in PLs one can exploit the above mentioned time scale difference, Zadeh's expansion for the entire channel response is useful in a more general set of cases. In fact, it allows for the development of adaptive techniques for the estimation of the harmonic responses of the channel (see Sect. \ref{sec.ZadehEstimation}).

\section{Lifted Representation of PL Channels} \label{sec.LiftedForm}

Let us assume that a link is constituted of $N$ cascaded sections and let $h^{(1,\cdots,N)}(t,\tau)$ be its TV overall channel impulse response. The signal at the output of the channel can be written as:
\begin{equation}\label{eq.final2_1}
v_{out}(t)=\int_{-\infty}^{\infty} v_s(\tau) h^{(1,\cdots,N)}(t,\tau) d\tau.
\end{equation}
In DT, the equivalent relationship is:
\begin{equation}\label{eq.final3}
    v_{out}[k]=\sum_{n=-\infty}^{\infty} v_s[n] h^{(1,\cdots,N)}[k,k-n],
\end{equation}
where the expression of $h^{(1,\cdots,N)}[k,l]$ is given in \eqref{eq.tv-rel11} at the top of the next page.
\begin{figure*}[!t]
\begin{equation}\label{eq.tv-rel11}
h^{(1,\cdots,N)}[k,l]=\int_{-\infty}^{\infty} \int_{-\infty}^{\infty} h^{(1,\cdots,N)}(kT_s-\xi,\tau) p(lT_s-\tau-\xi) p'(\xi) d\tau d\xi
\end{equation}
\hrulefill
\vspace*{4pt}
\end{figure*}
$p(t)$ is the transmit pulse shaping filter, and $p'(t)$ is the receive filter and $h^{(1,\cdots,N)}(t,\tau)$ is the baseband or baseband equivalent response of the LTV channel. Now, omitting superscript $(1,\cdots,N)$ for brevity, our objective is to cast \eqref{eq.final3} in the following vector matrix model \cite{ScaGiaBar2002}:
\begin{equation}\label{eq.ABCD16}
  \bv_{out}[i] = {\bH}_{i,0}  \bv_{s}[i]+{\bH}_{i,1} \bv_{s}[i-1].
\end{equation}
This is possible by {\it lifting} the streams $v_{out}[k]$ and $v_s[n]$ into blocks as follows: $\{\bv_s[i]\}_{k}$ $\triangleq$ $v_{s}[iP+k]$, $\{\bv_{out}[i]\}_{k}$ $\triangleq$ $v_{out}[iP+k]$, and by introducing the following two matrices $(k,n=0,\cdots,P-1)$:
%

\begin{IEEEeqnarray}{rCl} \label{eq.H12}
  \{ {\bH}_{i,0} \}_{k,n} &=& h[iP+k,k-n]\\
  \{ {\bH}_{i,1} \}_{k,n} &=& h[iP+k,P+k-n].
\end{IEEEeqnarray}
%
Matrices ${\bH}_{i,0}$ and ${\bH}_{i,1}$ are banded and their structure is shown in Fig. \ref{fig.H1}.
The underlying assumption for the representation in \eqref{eq.ABCD16} is that the channel $h[k,l]$ has finite memory $L$ smaller that the block length $P$, i.e. $P>L$ and $h[k,l]=0$ for $l<0$ and $l>L$. This allows us to limit IBI only to two consecutive terms. If this condition were not met, then the output block $\bv_{out}[i]$ would be a function of infinite IBI terms.

Since only the top right corner of ${\bH}_{i,1}$ is non zero, having $L$ zeros in the last $L$ entries of every size-$P$ transmitted block allows us to eliminate completely IBI thus yielding to simplified expressions (TZ case). In fact, posing $\bv_{s}[i]=(\underline{\bv}_{s}[i],0,\ldots,0)$ allows us to replace \eqref{eq.ABCD16} with an IBI free counterpart:
\begin{equation}\label{eq.ABCD17}
  \bv_{out}[i] = {\bH}_{i,0}  \bv_{s}[i]={\bbH}_i \underline{\bv}_{s}[i].
\end{equation}

\insertonefig{8.8cm}{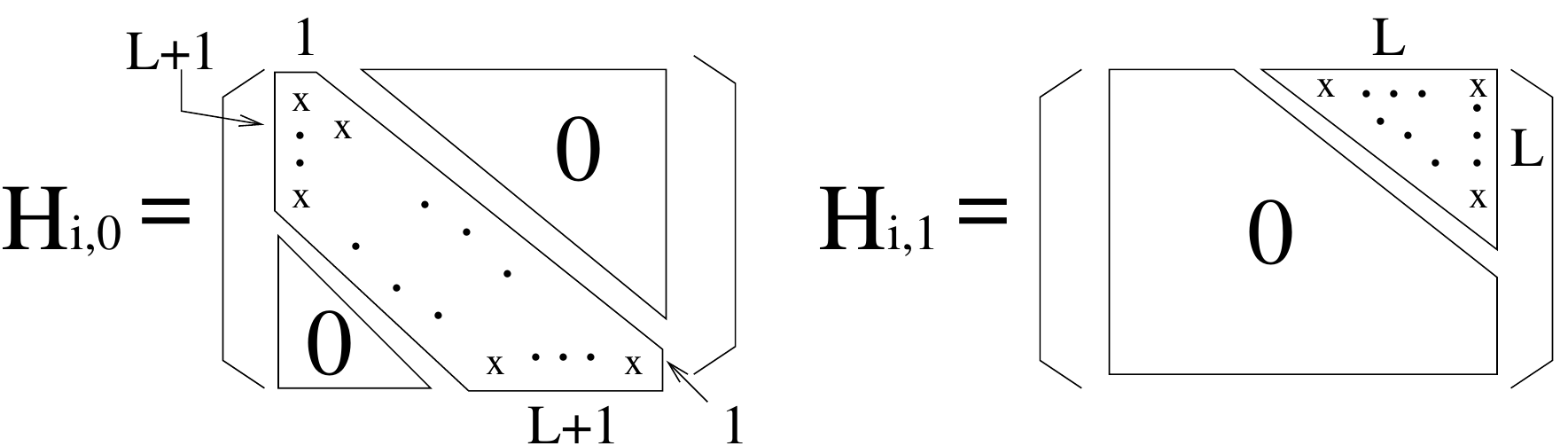}{\label{fig.H1}}
{Time-varying block channel matrices have a lower-banded (left) and an upper triangular (right) structure of order $L$. The elements of these matrices are given in \eqref{eq.H12}.}

\begin{remark}\label{remark.numerical-instability}
Even though ${\bH}_{i,0}$ in \eqref{eq.ABCD17} is a lower triangular matrix and, therefore, it is invertible, its condition number is significantly affected by the channel response. Such a matrix is well conditioned only when the channel is minimum phase.  Interestingly, its tall counterpart ${\bbH}_i$, obtained by removing the last $L$ columns of $\bH_i$, is typically well conditioned and its properties are discussed in detail in \cite{ScaBarGian1998}. In general, in solving for the input, one should invert the over-determined system ${\bbH}\underline{\bx}[i]$ rather than inverting ${\bH}_{0}$, to avoid numerical instability.
\end{remark}

\subsection{Estimation of Zadeh's Harmonic Impulse Responses} \label{sec.ZadehEstimation}

Let us assume that $f_0\ll W$, where $W$ is the bandwidth of the receive filter $p'(\tau)$, and let $h_m[l]\triangleq \int_{-\infty}^{\infty} \int_{-\infty}^{\infty} h_m(\tau) p(lT_s-\tau-\xi) p'(\xi) d\tau d\xi$, where $h_m(\tau)$ are the baseband or baseband equivalent harmonic responses in \eqref{eq.ctmodel-chanFS2}. Note that, given the receiver filtering, any component of the LPTV channel outside the receive filter will be filtered out.
Introducing the following matrices for $k,n=0,\cdots,P-1$, and $l=0,1$:
\begin{eqnarray}
\{{ \hat \bH}_{m,l} \}_{k,n} &\triangleq& h_{m}[lP+k-n] \\
\{{ \bf \Omega}_{m} \}_{k,n} &\triangleq& e^{j2\pi m f_0 k T_s}\delta[n-k],
\end{eqnarray}
Zadeh expansion for the channel matrix in block form is as follows:
%
\begin{IEEEeqnarray}{rCl} \label{eq.Zadehblock}
\{ { \bH}_{i,l} \}_{k,n} &=& \sum_{m} h_{m}[lP+k-n] e^{j2\pi m f_0 (iP+k) T_s}\\
                         &=& \sum_{m} e^{j2\pi m f_0 i P T_s}\{{ \bf \Omega}_{m} { \hat \bH}_{m,l} \}_{k,n}.
\end{IEEEeqnarray}
%
In the case of TZ, assuming that the $h_{m}[l]$ have finite memory $L$, denoting by $\bh_m$ the vectors of their non zero coefficients, and by $\{{ \bbH}_{i} \}_{k,n}=\sum_{m} h_{m}[k-n] e^{j2\pi f_0 T_s m (iP+k)}$ for $k\in[0,P-1], n\in[0,P-L-1]$ we can analogously remove the IBI and have:
\begin{eqnarray}\label{eq.ZadehblockTZ}
  \bv_{out}[i] &=& \sum_{m} e^{j2\pi m f_0 i P T_s}{ \bf \Omega}_{m} { \hat \bbH}_{m} \underline{\bv}_s[i] \\
               &=& \sum_{m} e^{j2\pi m f_0 i P T_s }{ \bf \Omega}_{m} {\bf \Phi}(\underline{\bv}_s[i])\bh_m,
\end{eqnarray}
where the $P\times L$ Toeplitz matrix $ {\bf \Phi}(\underline{\bv}_s[i])$ has first column $\bv_s[i]$ and first row $(v_s(iP),0,\ldots,0)$.

Assuming that the response can be well approximated with a limited number of harmonics $m\in[-M,M]$, stacking all the kernels in $\bh=(\bh^T_{-M},\ldots,\bh^T_{M})^T$ and forming the matrix ${\bf \Psi}(\underline{\bv}_s[i])=(e^{-j2\pi M f_0 i P T_s}{ \bf \Omega}_{-M} {\bf \Phi}(\underline{\bv}_s[i]),\ldots, e^{j2\pi M f_0 i P T_s}{ \bf \Omega}_{M} {\bf \Phi}(\underline{\bv}_s[i]))$, this means that:
\begin{equation}
  \bv_{out}[i]={\bf \Psi}(\underline{\bv}_s[i])\bh
\end{equation}
and a unique MMSE estimate of Zadeh's harmonic impulse responses exists if the matrix is at least square ($P\geq (2M+1)L$) and invertible. The MMSE estimate of $\bh$ is then:
\begin{equation} \label{ZadehEstimation}
\hat{\bh}={\bf \Psi}^{\dagger}(\underline{\bv}_s[i])  \bv_{out}[i].
\end{equation}
Note that estimate \eqref{ZadehEstimation} is more accurate than the one proposed in equation (4) of \cite{CanCorDie2006} which is an ML estimate that neglects noise and also requires the use of a Fourier basis.

We have just shown that \eqref{ZadehEstimation} allows to estimate the LPTV channel from the sampled observations at the receiver. The error in the estimated channel response can be readily quantified after recognizing that it includes three terms: 1) the residual MSE due to noise in the observations which is given by $trace( ({\bf \Psi}^H(\underline{\bv}_s[i]){\mathbf{R}_n}{\bf \Psi}(\underline{\bv}_s[i]))^{-1} )$, where $\mathbf{R}_n$ is the noise covariance matrix\footnote{~The noise covariance matrix $\mathbf{R}_n$ is seldom diagonal as noise in PLs is colored (see \cite{Big2003} and references therein).}; 2) the series truncation errors given in \eqref{eq.TruncationFourier} due to considering a limited number of harmonics $m\in[-M,M]$ and, last but not least, 3) the approximation of the responses with Finite Impulse Response (FIR) discrete time filters. The nature of this last approximation will be discussed and quantified in the next Section.

\subsection{On the Finite Support of the ABCD Kernels} \label{sec.LTIcase}
The time scale of the excitation given by the AC signal is very slow compared to both the duration of the responses of the LPTV 2PN as well as the Nyquist interval for transmission over PLs, given the broad bandwidth of the input signal itself. This, we argue, implies that not only the load equations can be linearized, but also that their response can be approximated resorting to the solution of the small signal equivalent circuit, as if it were stationary. Hence, it is possible to map the ABCD parameters in linear impulse responses. Based on \eqref{eq.ABCDdef} it is clear that the kernels $a(t,\tau), d(t,\tau)$, which are different from one only for an LTI section of cable, are always LTI impulse responses $a(\tau),d(\tau)$ (see Section \ref{sec.singlesection}) while, for series and shunt impedances, this yields:
\begin{eqnarray}
b(t,\tau)=\sum_{m=-\infty}^{+\infty} b_m(\tau)e^{j 2 \pi m f_0 t} \\
c(t,\tau)=\sum_{m=-\infty}^{+\infty} c_m(\tau)e^{j 2 \pi m f_0 t}.
\end{eqnarray}
The above expressions are TV once the relationship between the periodic AC bias and the small signal circuit parameters that determine their impedance is made explicit. Given the $Z(t,f)$ the Zadeh expansion terms are as follows:
%
\begin{IEEEeqnarray}{rCl}
b_m(\tau) &\approx& \frac{1}{T_0}\int_{-\infty}^{+\infty}\int_{0}^{T_0} Z(t,f)e^{-j2\pi m f_0 t }e^{j2\pi f \tau }dt df \IEEEeqnarraynumspace\\
c_m(\tau) &\approx& \frac{1}{T_0} \int_{-\infty}^{+\infty}\int_{0}^{T_0} Z^{-1}(t,f)e^{-j2\pi m f_0 t }e^{j2\pi f \tau }dt df \IEEEeqnarraynumspace
\end{IEEEeqnarray}
%
Note that the low pass equivalent kernel, which are of greater interest for PLC, can be obtained by simply replacing $Z(t,f)$ with $Z(t,f+f_c)$, i.e. down-shifting the expression of the impedance by the carrier frequency $f_c$.

At this point, the question we want to address is whether an analytical path towards generating the ${\bH}_{i,j}$ on the basis of the TL formalism exists or not. Specifically, the issue at hand is whether or not one can resort to a DT lifted representation of the TV kernels in the TL equations \eqref{eq.tvconv11b} to get to \eqref{eq.ABCD16}.

As mentioned in Sect. \ref{sec.LiftedForm}, the underlying assumption for exploiting lifting is that the kernels must have finite memory, and the objective of this section is to verify it.

Based on the separation of time scales discussed earlier, in our analysis we treat each section of the PL link as LTI. When representing LTV loads, this is exact for the parameters A and D and it is valid for parameters B and C when the separation of time scales holds. When representing sections of cable or bridged taps, this will also be exact as they obviously represent LTI channels (assumption of uniform TLs).

For the LTI case, one can rewrite \eqref{eq.IO5} expressing the electrical quantities on Port 2 (output) as the result of a convolution between the electrical quantities on Port 1 (input) and LTI filtering kernels:
%
\begin{IEEEeqnarray}{rCl}
v_{out}(t) &=& \int_{-\infty}^{\infty} v_{in}(\tau) d(t-\tau) d\tau \\
    && -\: \int_{-\infty}^{\infty} i_{in}(\tau) b(t-\tau) d\tau \nonumber \\
i_{out}(t) &=& -\int_{-\infty}^{\infty} v_{in}(\tau) c(t-\tau) d\tau \\
    && +\: \int_{-\infty}^{\infty} i_{in}(\tau) a(t-\tau) d\tau \label{eq.tvconv11a}
\end{IEEEeqnarray}
%
where $a(t)$, $b(t)$, $c(t)$, and $d(t)$ are the impulse responses of the 2PN filtering kernels and they are obtained by Inverse FT (IFT) of $A(f)$, $B(f)$, $C(f)$, and $D(f)$, respectively. Relationship \eqref{eq.tvconv11a} holds for both the cases where the filtering kernels represent a single 2PN of the kind of \eqref{eq.ABCDdef} or the overall TM of a cascade of 2PNs.

To verify that $a(t)$, $b(t)$, $c(t)$, and $d(t)$ have finite memory, we study the following three cases: a single section of cable, single and cascaded shunt impedances, and single and cascaded series impedances. Without loss of generality, we resort to the baseband kernel model.

\subsubsection{Single Section of Cable}\label{sec.singlesection}
The forward ABCD parameters have the closed form expression in \eqref{eq.ABCDdef}. A single section of cable acts on the transmitted signal as a low-pass filter with decreasing cut-off frequency as the length of the section increases; thus, it can be modeled as an IIR-like filter and, the longer the cable, the longer the impulse response and the more the IIR-like behavior will be. This can be verified analytically by noting that filters $1/A(f)$ and $1/D(f)$ have the following known FT pair \cite{Bracewell1965}:
\begin{equation}
\mbox{sech}(at)   \Leftrightarrow   \frac{\pi}{a} \mbox{sech}\left(\frac{\pi^2}{a} f\right),
\end{equation}
and $\lim_{t\rightarrow \infty} \text{sech$(at)$}=0$. It is immediate to recognize that the DT version of the inverse filters of $A(f)$ and $D(f)$ are likely to be well approximated by FIR responses while the impulse responses $a(t)$ and $d(t)$ are better approximated by IIR filters and thus would result in IBI from infinite blocks. However, truncating the IIR response leads to a small error when the section of cable is not too long. This can be verified looking at Fig. \ref{fig.Kernelat} where we can see that the filtering kernel $a(t)=$IFT$\{A(f)\}$ for a single section of telephone AWG 24 cable is FIR-like until lengths of 1.5 kft but starts becoming IIR-like around 2 kft. For example, at a length of $1.5$ kft we have ascertained that the energy of $a(t)$ truncated after $0.75\mu s$ is equal to around $98.5\%$ of the energy of the non-truncated response. We also note that $a(t)$ is equal to the impulse response of the link for the lengths shown in Fig. \ref{fig.Kernelat}.(a). This fact is not surprising as, when the second port is unterminated, the output of the filter $A(f)$ is equal to the voltage on the second port and the ratio of the voltages at the two ports is equal to the impulse response for an ideal source.

\inserttwofigsV{8.8cm}{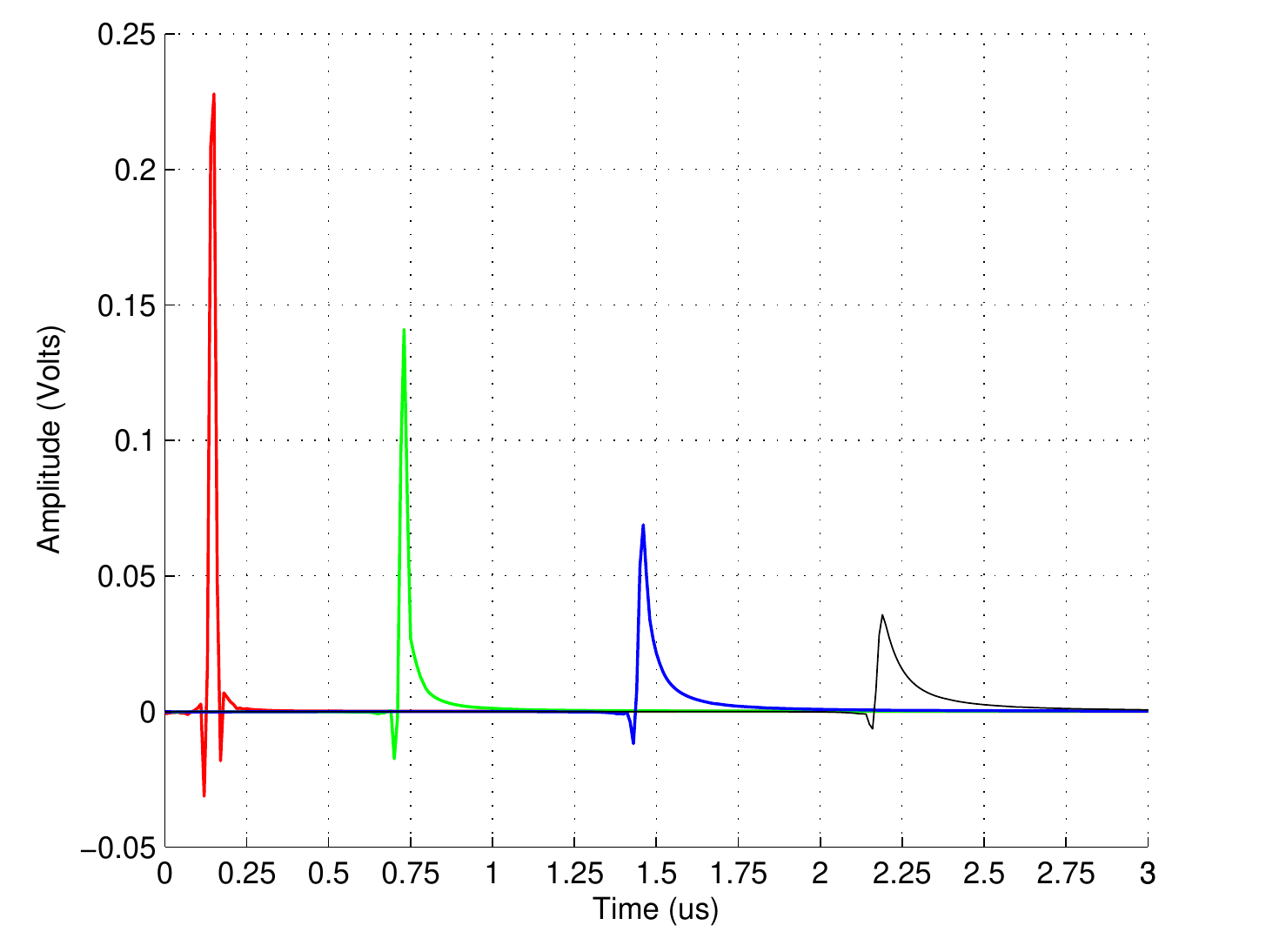}{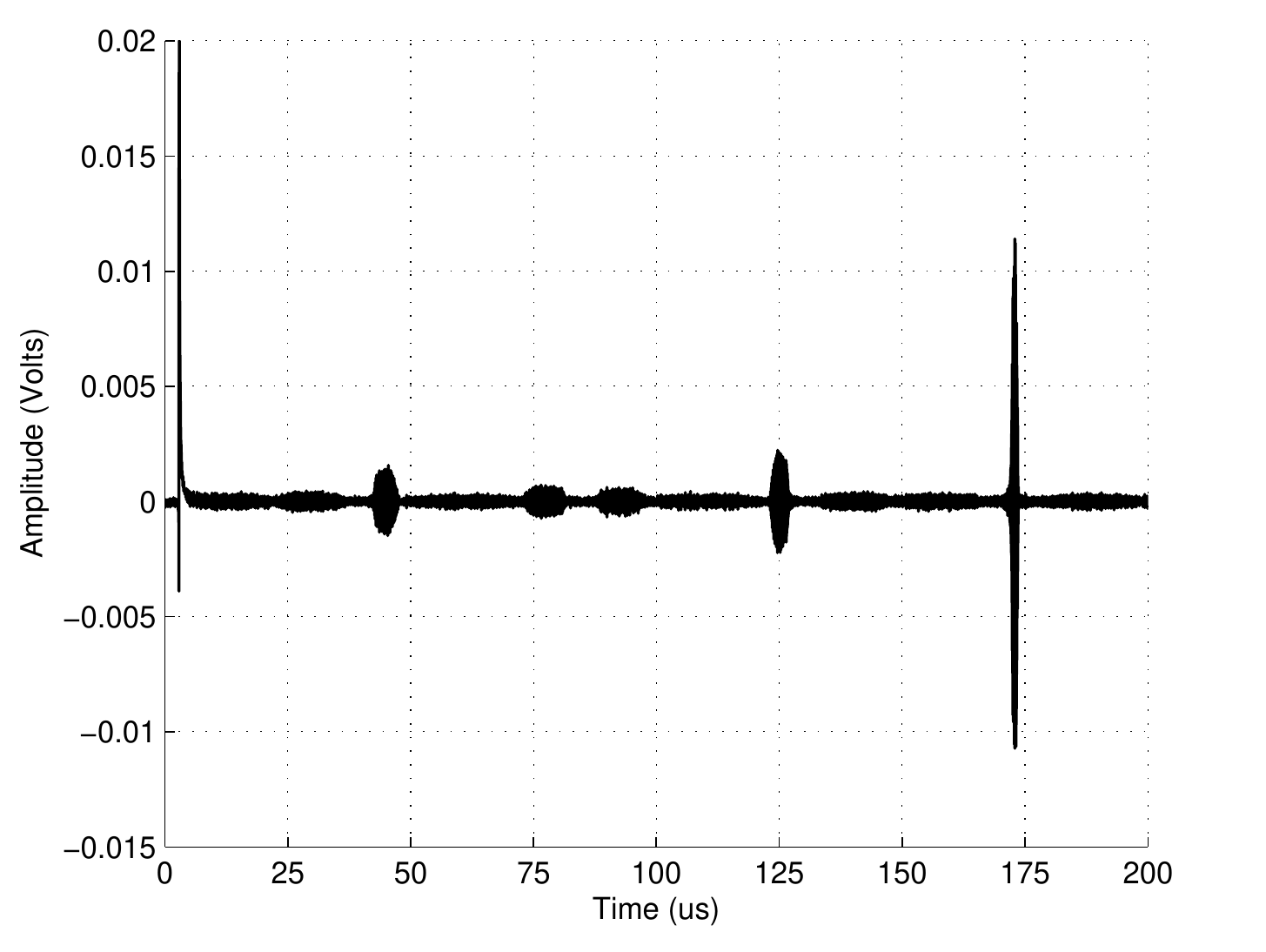}{\label{fig.Kernelat}}
{Plot of the kernel $a(t)=$IFT$\{A(f)\}$ for the case of single section of an AWG24 cable of length $l$ feet, on a band between DC and 30 MHz after raised cosine filtering with roll off equal to 0.5. (a) $l=100, 500, 1000,$ and $1,500$, from left to right; (b) $l=2,000$.}

For the case of outdoor topologies, one can split the longer cable sections in multiple sub-section of smaller lengths, find the DT lifted version of these sub-sections and then consider the cascade of these multiple sections to model the whole cable. In the FD, this is easily done by using the Chain Rule. In Sect. \ref{sec.ChainRule} we will show that it is always possible to handle tandem connection of 2PNs even in the DT lifted form; furthermore, we show how a DT lifted equivalent of the FD Chair Rule exists under the TZs assumption.

As an alternative for the case of cable sections of longer length, we can also introduce new kernels rewriting \eqref{eq.IO} as follows ($A(f)$ and $D(f)$ are always non-zero):
\begin{eqnarray}
V_{out}(f)  &=& \frac{1}{A(f)} V_{in}(f) + \frac{-B(f)}{A(f)} I_{out}(f)  \nonumber \\
I_{out}(f)  &=& \frac{1}{D(f)} I_{in}(f) + \frac{-C(f)}{D(f)}  V_{out}(f) \label{eq.IO5b}
\end{eqnarray}
The new filters $\alpha(t)$, $\beta(t)$, $\gamma(t)$, and $\zeta(t)$ have impulse responses that are obtained by IFT of $1/A(f)$, $-B(f)/A(f)$, $1/D(f)$, and $-C(f)/D(f)$, respectively. The transfer functions of these new kernels are of the kind sech$(f)$ and $\tanh(f)$. Since the following FT relationships hold \cite{Bracewell1965}:
\begin{equation}\label{eq.FTkernels}
\begin{cases}
\text{sech$(at)$}    &  \Leftrightarrow   \frac{\pi}{a} \text{sech($\frac{\pi^2}{a} f$)} \\
\text{cosech$(at)$}  &  \Leftrightarrow   \frac{\pi}{a} \tanh(\frac{\pi^2}{a} f)
\end{cases}
\end{equation}
and we also have that $\lim_{t\rightarrow \infty} \text{cosech$(at)$}=0$, these kernels decay exponentially. The filtering kernels $\alpha(t)$, $\beta(t)$, $\gamma(t)$, and $\zeta(t)$ for the case of a 2 kft AWG 24 cable are shown in Fig. \ref{fig.KernelsAlpha}.(a), and $\alpha(t)$ is compared to the impulse response $h(t)$ of the link in Fig. \ref{fig.KernelsAlpha}.(b). Besides verifying that the filtering kernels are well approximated by FIR filers, we also note that $\alpha(t)$ coincides with $h(t)$ for the first $9\mu s$ (apart from a scaling factor) and then is followed by echoes of alternating sign.

\inserttwofigsV{8.8cm}{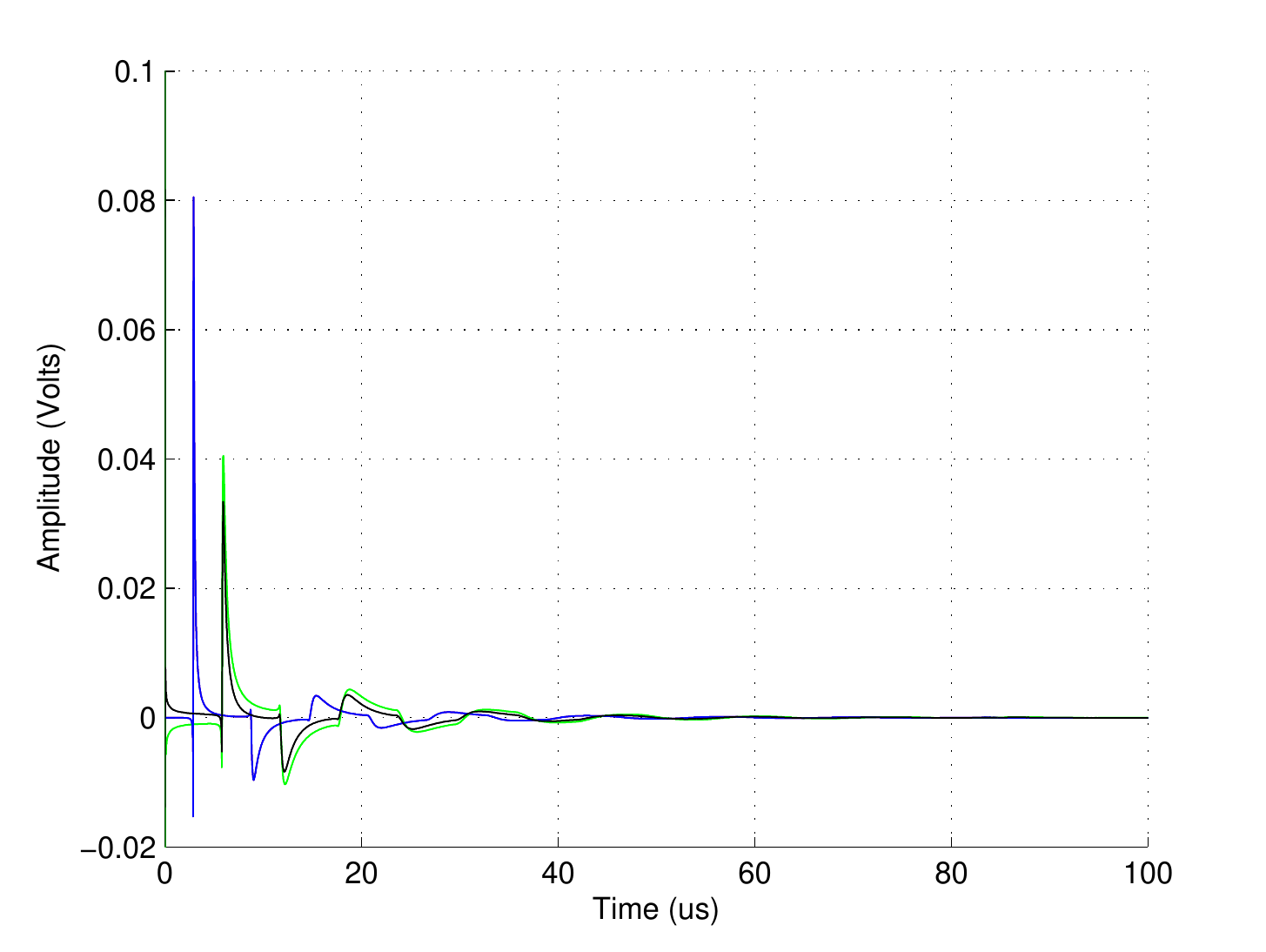}{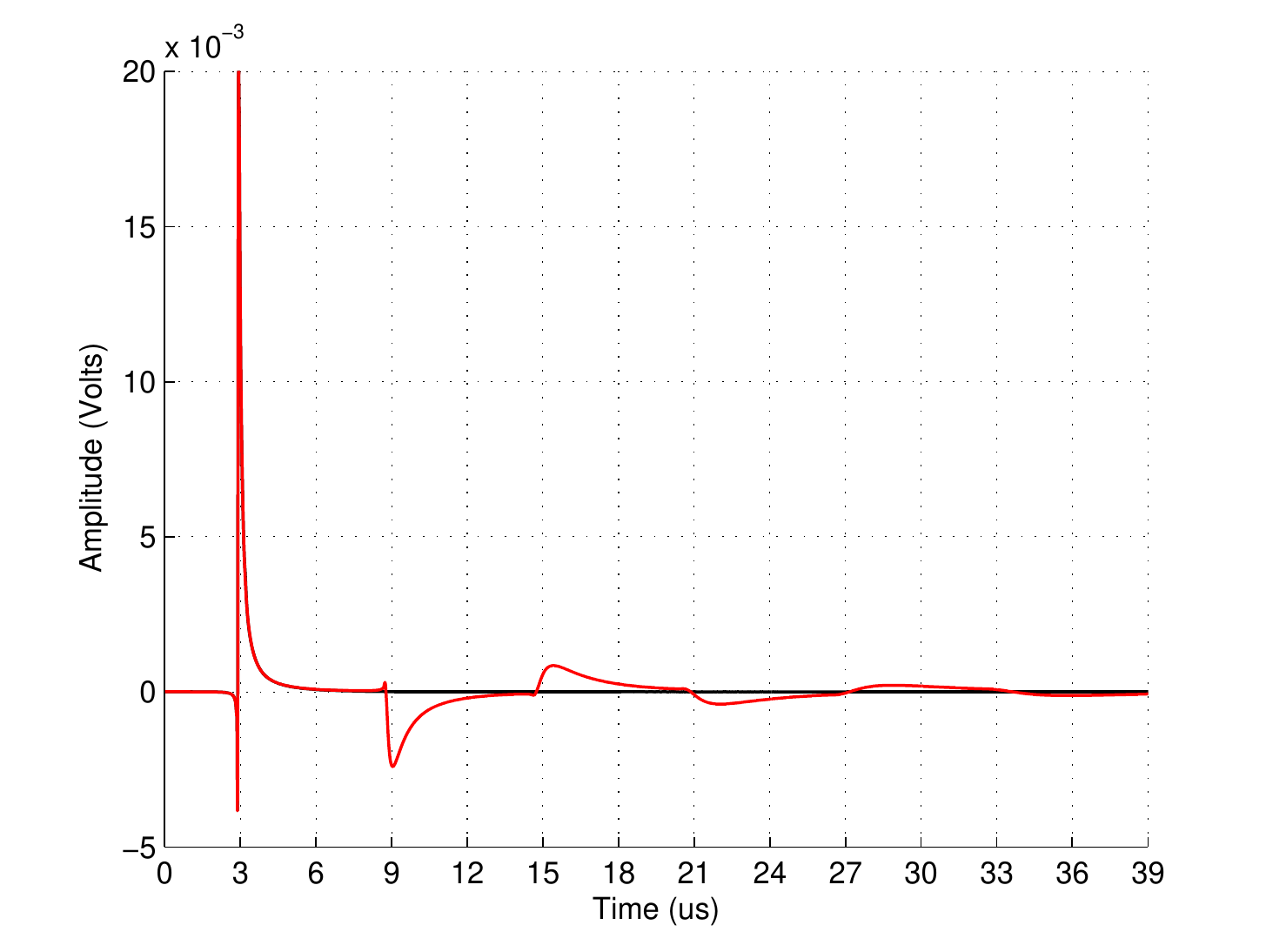}{\label{fig.KernelsAlpha}}
{Case of a 2,000 ft long AWG 24 cable. (a) Plot of the filtering kernels $\alpha(t)=\gamma(t)$, $\beta(t)/50$, and $150\zeta(t)$ of \eqref{eq.IO5b}. (b) Comparison between the impulse response $h(t)$ (black) and the kernel $4\alpha(t)$ (red).}

Again this is not surprising as the output of the filter $1/A(f)$ when the second port is unterminated is equal to the voltage on the first port and, therefore, these alternating echoes are the successive reflections between the unterminated second port and the first port. In fact, echoes are spaced around $6\mu s$ from each other which corresponds to a signal traveling 4,000 feet, the roundtrip of the length of the cable. This result is in agreement with \cite{GalWar02} where a model for the echoes bouncing between discontinuities was developed for the PH case. Although infinite echoes are theoretically present, only few echoes are non-negligible due to cable attenuation. The fact that IFT$\{A(f)\}$  and IFT$\{1/A(f)\}$ can well approximate the impulse response of a TL link has never been noted and it is reported here for the first time.

\subsubsection{Single and Cascaded Shunt Impedances}

Shunt impedances model appliances plugged into an outlet as well as bridged taps \cite{Wer1991}. In the vast majority of cases, a bridged tap is composed of a single section of cable terminated on a constant impedance $Z_t$. For shunt impedances, the kernels are $a(t)=d(t)=\delta(t)$, $b(t)=0$, and $c(t)=$IFT$\{1/Z_{in}(f)\}$ (see \eqref{eq.ABCDdef}) where:
\begin{equation}
Z_{in}(f) = \frac{A(f) \cdot Z_t(f)+B(f)}{C(f) \cdot Z_t(f)+D(f)}
\end{equation}
The above expression can be simplified further in the HF region as follows (see Fig. \ref{fig.GammaZo}):
\begin{equation} \label{eq.BTsimpleCf}
C(f) = \frac{Z_t(f)+Z_o^\infty}{Z_o^\infty(Z_t(f)+Z_o^\infty)}
\end{equation}
where we have exploited $\lim_{x\rightarrow \infty} \cosh(x)\approx \lim_{x\rightarrow \infty} \sinh(x)$ and posed $\lim_{f\rightarrow \infty} Z_o(f)=Z_o^\infty$. An IH bridged tap is often terminated with an open, a short,  a resistor, or a small capacitor, so that in the HF band and above $C(f)$ is constant and we can write $c(t) \approx k \delta(t)$ and its kernels have no memory at all. The same considerations apply to $N$ cascaded shunt impedances since their combined TM has the same form of \eqref{eq.ABCDdef} but with $C=\sum_{i=1}^{N} 1/Z^{(i)}$, where $Z^{(i)}$ is the $i$-th shunt impedance in the cascade.

\subsubsection{Single and Cascaded Series Impedances}
We have that $a(t)=d(t)=\delta(t)$, $c(t)=0$, and $b(t)$ is equal to the IFT of the series impedance (see \eqref{eq.ABCDdef}). As for the shunt impedance case, it is easy to verify that the TM of the cascade of $N$ series  impedances still has the same form of \eqref{eq.ABCDdef} but with $B=\sum_{i=1}^{N} Z^{(i)}$, where $Z^{(i)}$ is the $i$-th series impedance in the cascade. We assume here that $b(t)$ is always FIR-like in all cases of practical interest.

\begin{remark}
The considerations made here for the LTI case are novel, and the existence of a (approximately) finite support for the LTI kernels is reported here for the first time. Since LTV TLs are composed by LTI sections (wires) and LTV sections (shunt/series impedances), the capability of handling both cases under the same formalism is of interest.
\end{remark}

\subsection{The TL Input-Output Relationship in Lifted Form}\label{sec.IO-TL}
To express the TL input-output as in \eqref{eq.ABCD16}, we will first tackle the problem of expressing in DT lifted form the continuous TD input-output relationships in \eqref{eq.tvconv11b}. This, in turn, will give us a DT equivalent of the Chain Rule to handle the cascade of 2PNs. The DT equivalent of \eqref{eq.tvconv11b} can be written as follows:
\begin{IEEEeqnarray}{rCl}
v_{out}[k] &=& \sum_{n=-\infty}^{\infty} v_{in}[n]d[k,k-n] \\
           && -\: \sum_{n=-\infty}^{\infty} i_{in}[n]b[k,k-n] \nonumber
\end{IEEEeqnarray}
\begin{IEEEeqnarray}{rCl}
i_{out}[k] &=& -\sum_{n=-\infty}^{\infty} v_{in}[n] c[k,k-n] \\
           && +\: \sum_{n=-\infty}^{\infty} i_{in}[n]a[k,k-n] \nonumber  \label{eq.tvconv12a}
\end{IEEEeqnarray}
where the LTV DT kernels above are defined as in \eqref{eq.tv-rel11}. Other relationships that will be useful for finding the DT input-output relationship of a TL link are based on Kirchoff laws:
\begin{eqnarray}
  v_{in}[k] &=& v_s[k]- z_0^{(s)} i_{in}[k] \label{eq.tvconv13a} \\
  v_{out}[k] &=&  z_0^{(L)} i_{out} [k]     \label{eq.tvconv13b}
\end{eqnarray}
Since we have ascertained the existence of either exact or approximated FIR filtering kernels that have finite memory $L$, we can rewrite the previous DT equations in lifted form for $P>L$:
%
\begin{IEEEeqnarray}{rCl} 
  \bv_{out}[i] &=& {\bD}_{i,0}  \bv_{in}[i] + {\bD}_{i,1}  \bv_{in}[i-1] \\
    && -\: {\bB}_{i,0}  \bi_{in}[i] - {\bB}_{i,1}  \bi_{in}[i-1] \label{eq.ABCD15a} \nonumber \\
  \bi_{out}[i] &=& -{\bC}_{i,0}  \bv_{in}[i] - {\bC}_{i,1}  \bv_{in}[i-1] \\
    && +\: {\bA}_{i,0}  \bi_{in}[i] + {\bA}_{i,1}  \bi_{in}[i-1] \label{eq.ABCD15b} \nonumber \\
  \bv_{in}[i]&=& \bv_{s}[i]-z^{(s)}_{0} \bi_{in}[i]  \label{eq.ABCD15c} \\
  \bv_{out}[i]&=& z^{(L)}_{0} \bi_{out}[i] \label{eq.ABCD15d}
\end{IEEEeqnarray}
%
where we have posed: $\{\bv_s[i]\}_{k}$ $\triangleq$ $v_{s}[iP+k]$, $\{\bv_{out}[i]\}_{k}$ $\triangleq$ $v_{out}[iP+k]$,
$\{\bv_{in}[i]\}_{k}$ $\triangleq$ $v_{in}[iP+k]$, $\{\bi_{out}[i]\}_{k}$ $\triangleq$ $i_{out}[iP+k]$  and $\{\bi_{in}[i]\}_{k}$ $\triangleq$ $i_{in}[iP+k]$.

The $P$-by-$P$ matrices ${\bA}_{i,\star}$, ${\bB}_{i,\star}$, ${\bC}_{i,\star}$ and ${\bD}_{i,\star}$ are defined as in \eqref{eq.H12}, e.g. $(k,n=0,\cdots,P-1) $:
\begin{eqnarray} \label{eq.LiftedDefs}
   \{ {\bA}_{i,i-j} \}_{k,n}=a[iP+k,(i-j)P+k-n] \hspace{0.5cm} 
\end{eqnarray}
On the basis of the considerations made at the end of Sect. \ref{sec.model-TVchannel-LTV}, we can give explicit expressions to some of the above matrices as shown in Table \ref{tab.Simplifications}.

Assuming the system is at rest before the first block of the excitation $\bv_s[1]$ is transmitted, one can use the four equations \eqref{eq.ABCD15a}-\eqref{eq.ABCD15d} to solve for the four unknown blocks $\bv_{out}[1]$, $\bi_{out}[1]$, $\bv_{in}[1]$, and $\bi_{in}[1]$ by posing $\bv_{in}[0]=0$ and $\bi_{in}[0]=0$. At the next transmitted i-th blocks ($i>1$), one must also carry the IBI terms as known vectors in the same four equations and solve again for the same four unknown blocks. At every step, a system of $4P$ unknowns and $4P$ equation must be solved.

If TZs are applied to the transmit signal $\bv_s[i]$, one can find the solution of the system without IBI. In fact, after eliminating $\bv_{in}[i]$ using \eqref{eq.ABCD15c}, we can rewrite \eqref{eq.ABCD15a}-\eqref{eq.ABCD15d} as follows:
%
\begin{IEEEeqnarray}{rCl} \label{eq.FinalDT1}
\left[
  \begin{array}{c}
    \bv_{out}[i] \\
    \bi_{out}[i] \\
  \end{array}
\right] =  \left[ \!\!
     \begin{array}{cc}
       \bD_{i,0} & -(z_0^{(s)} \bD_{i,0}+\bB_{i,0}) \\
      -\bC_{i,0} &  (z_0^{(s)} \bC_{i,0}+\bA_{i,0}) \\
     \end{array}
  \!\!  \right] \left[ \!\!  \begin{array}{c}
    \bv_{s}[i] \\
    \bi_{in}[i] \\
  \end{array}
 \!\!   \right] \IEEEeqnarraynumspace
\end{IEEEeqnarray}
%
Now, by exploiting also \eqref{eq.ABCD15d} we can write:
%
\begin{equation} \label{eq.FinalDT2}
[
  \begin{array}{c}
    \bI   -z_0^{(L)} \bI \\
  \end{array}
]
  \left[\!\!
     \begin{array}{cc}
       \bD_{i,0} & -(z_0^{(s)} \bD_{i,0}+\bB_{i,0}) \\
      -\bC_{i,0} &  (z_0^{(s)} \bC_{i,0}+\bA_{i,0}) \\
     \end{array}
  \!\!  \right] \left[ \!\! \begin{array}{c}
    \bv_{s}[i] \\
    \bi_{in}[i] \\
  \end{array}
   \!\!\right]
   =
   \bzero 
\end{equation}
and solving for $\bi_{in}[i]$ we obtain \eqref{eq.FinalDT3} the following expressionL:
\begin{IEEEeqnarray}{rCl} \label{eq.FinalDT3}
  \bi_{in}[i] &=& \bXi_{i,0}^{\dagger} (\bD_{i,0}+z_0^{(L)}\bC_{i,0}) \bv_s[i]
\end{IEEEeqnarray}
where matrix $\bXi_{i,0}^{\dagger}$ is defined as
\begin{equation}\label{eq.FinalDT3bis}
  \bXi_{i,0}^{\dagger} = (z_0^{(s)}\bbD_{i,0} + \bbB_{i,0} + z_0^{(s)}z_0^{(L)}\bbC_{i,0} + z_0^{(L)}\bbA_{i,0})^{\dagger} 
\end{equation}
In the above equation we have taken into account the numerical instabilities that may arise when inverting lower banded matrices as mentioned in Remark \ref{remark.numerical-instability}. Thus, rather than inverting matrices $\bA_{i,0}$, $\bB_{i,0}$, $\bC_{i,0}$, and $\bD_{i,0}$ we have resorted to t  heir tall counterparts $\bbA_{i,0}$, $\bbB_{i,0}$, $\bbC_{i,0}$, and $\bbD_{i,0}$, respectively. We also point out that $\bXi_{i,0}$ resembles very closely the denominator of equation \eqref{eq.Hf} which yields the channel transfer function in the FD.

Finally, substituting the expression for $\bi_{in}[i]$ in \eqref{eq.FinalDT1}, we obtain the input-output expression given in \eqref{eq.FinalDTio} shown at the top of the next page, where in the last row we have replaced the $P\times 1$ input signal vector $\bv_s[i]$ with TZs with the $(P-L)\times 1$ vector $\underline{\bv}_s[i]$ defined as $\bv[i]=(\underline{\bv}_s^T[i],0,\ldots,0)^T$.

\begin{figure*}[!t]
\begin{IEEEeqnarray}{rCl} \label{eq.FinalDTio}
  \bv_{out}[i] = [\bbD_{i,0} - (z_0^{(s)} \bbD_{i,0}+\bbB_{i,0})\bXi_{i,0}^{\dagger} (\bbD_{i,0}+z_0^{(L)}\bbC_{i,0})] \underline{\bv}_s[i]
  \IEEEeqnarraynumspace
\end{IEEEeqnarray}
\hrulefill
\vspace*{4pt}
\end{figure*}

\section{DT Lifted Chain Rule} \label{sec.ChainRule}
It is of particular interest to check whether it is possible to extend the ABCD Chain Rule defined in the phasor domain to the DT lifted case. Let us start by expressing the input output-relationship of a 2PN in the following matrix form:
\begin{equation}\label{eq.ChainRuleDT}
\left[
  \begin{array}{c}
    \bv_{out}[i] \\
    \bi_{out}[i] \\
  \end{array}
\right]
=  \left[
     \begin{array}{cc}
       \overline{\bD}_i & -\overline{\bB}_i \\
       -\overline{\bC}_i & \overline{\bA}_i \\
     \end{array}
   \right] \left[  \begin{array}{c}
    \overline{\bv}_{in}[i] \\
    \overline{\bi}_{in}[i] \\
  \end{array}
   \right]
\end{equation}
where
\begin{equation}\label{eq.ChainDef}
\left\{
  \begin{array}{c}
    \overline{\bA}_i  =[{\bA}_{i,0} \vdots {\bA}_{i,1}]  \\
    \overline{\bB}_i  =[{\bB}_{i,0} \vdots {\bB}_{i,1}] \\
    \overline{\bC}_i  =[{\bC}_{i,0} \vdots {\bC}_{i,1}] \\
    \overline{\bD}_i  =[{\bD}_{i,0} \vdots {\bD}_{i,1}] \\
  \end{array}
\right. \hspace{0.3cm} \mbox{and} \hspace{0.3cm}     \left\{
  \begin{array}{c}
    \overline{\bv}_{in}[i]  =\left[
                               \begin{array}{c}
                                 \bv_{in}[i] \\
                                 \bv_{in}[i-1]
                               \end{array}
                             \right] \\
                             \\
    \overline{\bi}_{in}[i]  =\left[
                               \begin{array}{c}
                                 \bi_{in}[i] \\
                                 \bi_{in}[i-1]
                               \end{array}
                             \right]  \\
  \end{array}
\right.
\end{equation}
\noi We can then state the following Theorem.

\begin{theorem}\label{thm.ChainIBI}
\textit{The input-output relationship of the cascade of N 2PNs in DT lifted form is:}
\begin{equation}\label{eq.ABCD8}
\left[
  \begin{array}{c}
    \bv_{out}[i] \\
    \bi_{out}[i] \\
  \end{array}
\right]
=  \left[
     \begin{array}{cc}
       \overline{\bD}^{(1,\cdots,N)}_i & -\overline{\bB}^{(1,\cdots,N)}_i \\
       -\overline{\bC}^{(1,\cdots,N)}_i & \overline{\bA}^{(1,\cdots,N)}_i \\
     \end{array}
   \right] \left[  \begin{array}{c}
    \overline{\bv}_{in}[i] \\
    \overline{\bi}_{in}[i] \\
  \end{array}
   \right]
\end{equation}
\textit{where}
\begin{IEEEeqnarray}{rCl}
\overline{\bA}^{(1,\cdots,N)}_i  &=& [{\bA}^{(1,\cdots,N)}_{i,0} \vdots {\bA}^{(1,\cdots,N)}_{i,1}],\\ \overline{\bB}^{(1,\cdots,N)}_i  &=& [{\bB}^{(1,\cdots,N)}_{i,0} \vdots {\bB}^{(1,\cdots,N)}_{i,1}],
\end{IEEEeqnarray}
\begin{IEEEeqnarray}{rCl}
\overline{\bC}^{(1,\cdots,N)}_i  &=& [{\bC}^{(1,\cdots,N)}_{i,0} \vdots {\bC}^{(1,\cdots,N)}_{i,1}],\\ \overline{\bD}^{(1,\cdots,N)}_i  &=& [{\bD}^{(1,\cdots,N)}_{i,0} \vdots {\bD}^{(1,\cdots,N)}_{i,1}],
\end{IEEEeqnarray}
\noi \textit{and where the sub-blocks of these matrices can be computed recursively as follows:}

{
\allowdisplaybreaks
\begin{IEEEeqnarray*}{rCl}
  \bA^{(1,\cdots,k)}_{i,0} &=& \bC^{(k)}_{i,0}\bB^{(1,\cdots,k-1)}_{i,0}+\bA^{(k)}_{i,0}\bA^{(1,\cdots,k-1)}_{i,0} \\
  \bA^{(1,\cdots,k)}_{i,1} &=& \bC^{(k)}_{i,1}\bB^{(1,\cdots,k-1)}_{i-1,0}+\bC^{(k)}_{i,0}\bB^{(1,\cdots,k-1)}_{i,1} \\
                && +\: \bA^{(k)}_{i,1}\bA^{(1,\cdots,k-1)}_{i-1,0}+\bA^{(k)}_{i,0}\bA^{(1,\cdots,k-1)}_{i,1} \\
  \bB^{(1,\cdots,k)}_{i,0} &=& \bD^{(k)}_{i,0}\bB^{(1,\cdots,k-1)}_{i,0}+\bB^{(k)}_{i,0}\bA^{(1,\cdots,k-1)}_{i,0} \\
  \bB^{(1,\cdots,k)}_{i,1} &=& \bD^{(k)}_{i,1}\bB^{(1,\cdots,k-1)}_{i-1,0}+\bD^{(k)}_{i,0}\bB^{(1,\cdots,k-1)}_{i,1} \\
                && +\: \bB^{(k)}_{i,1}\bA^{(1,\cdots,k-1)}_{i-1,0}+\bB^{(k)}_{i,0}\bA^{(1,\cdots,k-1)}_{i,1} \\
  \bC^{(1,\cdots,N)}_{i,0} &=& \bC^{(k)}_{i,0}\bD^{(1,\cdots,k-1)}_{i,0}+\bA^{(k)}_{i,0}\bC^{(1,\cdots,k-1)}_{i,0} \\
  \bC^{(1,\cdots,k)}_{i,1} &=& \bC^{(k)}_{i,1}\bD^{(1,\cdots,k-1)}_{i-1,0}+\bC^{(k)}_{i,0}\bD^{(1,\cdots,k-1)}_{i,1} \\
                && +\: \bA^{(k)}_{i,1}\bC^{(1,\cdots,k-1)}_{i-1,0}+\bA^{(k)}_{i,0}\bC^{(1,\cdots,k-1)}_{i,1} \\
  \bD^{(1,\cdots,k)}_{i,0} &=& \bD^{(k)}_{i,0}\bD^{(1,\cdots,k-1)}_{i,0}+\bB^{(k)}_{i,0} \bC^{(1,\cdots,k-1)}_{i,0} \\
  \bD^{(1,\cdots,k)}_{i,1} &=& \bD^{(k)}_{i,1}\bD^{(1,\cdots,k-1)}_{i-1,0}+\bD^{(k)}_{i,0}\bD^{(1,\cdots,k-1)}_{i,1}\\
                && +\: \bB^{(k)}_{i,1}\bC^{(1,\cdots,k-1)}_{i-1,0}+\bB^{(k)}_{i,0}\bC^{(1,\cdots,k-1)}_{i,1}
\end{IEEEeqnarray*}
}
\end{theorem}

\vspace{0.25cm}

\begin{IEEEproof}
Using \eqref{eq.ABCD8} for $N=k$, the claim of the Theorem is obtained by recognizing that $\bv^{(1,\cdots,k-1)}_{out}[i]=\bv^{(k)}_{in}[i]$ and that $\bi^{(1,\cdots,k-1)}_{out}[i]=\bi^{(k)}_{in}[i]$ and that the product of two upper triangular matrices is always zero, i.e. ${\bH}^{(N)}_{i,1}{\bH}^{(N-1)}_{i,1}=\mathbf{0}$.
\end{IEEEproof}

\vspace{0.25cm}

\begin{remark}
The chain rule valid in the phasor domain allows us to multiply directly the transmission matrices to obtain the transmission matrix of the overall link. We recognize that this is not possible in the lifted form when IBI is present. However, if we add trailing zeros to the input signal $\bv_s[i]$, then we obtain the DT lifted equivalent of the Chain Rule of the phasor domain as shown in the following Corollary.
\end{remark}

\vspace{0.25cm}
\begin{corollary}
\textit{If the input signal has L trailing zeros, then the Chain Rule in the DT lifted form has the following expression for the cascade of N systems:}
%
\begin{IEEEeqnarray}{rCl} \label{eq.ABCD11}
\left[
  \begin{array}{c}
    \bv_{out}[i] \\
    \bi_{out}[i] \\
  \end{array}
\right] &=&
    \left[
      \begin{array}{cc}
        \bD^{(1,\cdots,N)}_{i,0}  & -\bB^{(1,\cdots,N)}_{i,0}   \\
        -\bC^{(1,\cdots,N)}_{i,0}  & \bA^{(1,\cdots,N)}_{i,0}   \\
      \end{array}
    \right]
    \left[
        \begin{array}{c}
        \bv_{in}[i] \\
        \bi_{in}[i] \\
        \end{array}
    \right] \IEEEeqnarraynumspace \\ \label{eq.ABCD11b}
    &=&
    \prod_{k=N}^{1}
    \left[
      \begin{array}{cc}
        \bD^{(k)}_{i,0}  & -\bB^{(k)}_{i,0}   \\
        -\bC^{(k)}_{i,0}  & \bA^{(k)}_{i,0}   \\
      \end{array}
    \right]
    \left[
        \begin{array}{c}
        \bv_{in}[i] \\
        \bi_{in}[i] \\
        \end{array}
   \right] \IEEEeqnarraynumspace
\end{IEEEeqnarray}
\textit{and the sub-blocks of matrix \eqref{eq.ABCD11} can be recursively expressed as follows:}
{
\allowdisplaybreaks
\begin{eqnarray}\label{eq.ABCD5}
  \bA^{(1,\cdots,k)}_{i,0} &=& \bC^{(k)}_{i,0}\bB^{(1,\cdots,k-1)}_{i,0}+\bA^{(k)}_{i,0}\bA^{(1,\cdots,k-1)}_{i,0} \\
  \bB^{(1,\cdots,k)}_{i,0} &=& \bD^{(k)}_{i,0}\bB^{(1,\cdots,k-1)}_{i,0}+\bB^{(k)}_{i,0}\bA^{(1,\cdots,k-1)}_{i,0} \\
  \bC^{(1,\cdots,k)}_{i,0} &=& \bC^{(k)}_{i,0}\bD^{(1,\cdots,k-1)}_{i,0}+\bA^{(k)}_{i,0}\bC^{(1,\cdots,k-1)}_{i,0} \\
  \bD^{(1,\cdots,k)}_{i,0} &=& \bD^{(k)}_{i,0}\bD^{(1,\cdots,k-1)}_{i,0}+\bB^{(k)}_{i,0}\bC^{(1,\cdots,k-1)}_{i,0}
\end{eqnarray}
}
\end{corollary}
\vspace{0.25cm}

\begin{IEEEproof}
The proof follows easily from from Theorem \ref{thm.ChainIBI}.
\end{IEEEproof}

\section{Conclusions} \label{sec.Conclusions}
We have proposed a DT equivalent model for TL-based communications channels which are usually modeled exclusively in the FD and only allow the treatment of the LTI case. The formalism adopted for our models allows us to address both the LTI and the LTV/LPTV cases which are of practical interest as they represent common home networking environments. Not only our method is suitable for more realistic simulations when the TL is LTV/LPTV, but also naturally leads to design of precoders and bit loading methods that are tailored to better cope with the periodically TV channel distortion since they directly relate the modulated symbols with the received data. These designs are going to be the subject of future work.

\begin{table*}[htbp]
  \centering
  \caption{Expression of the lifted version of the filtering kernels for some notable cases. Constant $k$ in row two can be calculated on the basis of \eqref{eq.BTsimpleCf} when $Z_t(f)$ is a constant impedance.}
  \label{tab.Simplifications}
  \begin{tabular}{lcccc}
    \hline\hline
     & $[\bA_{i,0}, \bA_{i,1}]$  &  $[\bB_{i,0}, \bB_{i,1}]$ & $[\bC_{i,0}, \bC_{i,1}]$ & $[\bD_{i,0}, \bD_{i,1}]$  \\ [0.5ex]
    \hline
    Shunt Impedance     & $[\bI, \; \bzero ]$ & $[\bzero, \; \bzero]$ & Load-dependent  & $[\bI, \;\bzero]$ \\
    Series Impedance    & $[\bI, \;\bzero ]$ & Load-dependent & $[\bzero, \; \bzero ]$  & $[\bI, \; \bzero ]$ \\
    Bridged Tap         & $[\bI, \; \bzero ]$ & $[\bzero, \; \bzero]$ & $[k\bI, \; \bzero]$ & $[\bI, \;\bzero]$ \\ [1ex]
    \hline
  \end{tabular}
\end{table*}

\appendices

\section*{Acknowledgment}
We want to thank Tae Eung Sung for long discussions on this subject and for bringing the authors together to complete the work.

\bibliographystyle{IEEEtran}
\bibliography{IEEEabrv,All_Papers,PLC_Papers}

\begin{IEEEbiography}[\addphoto{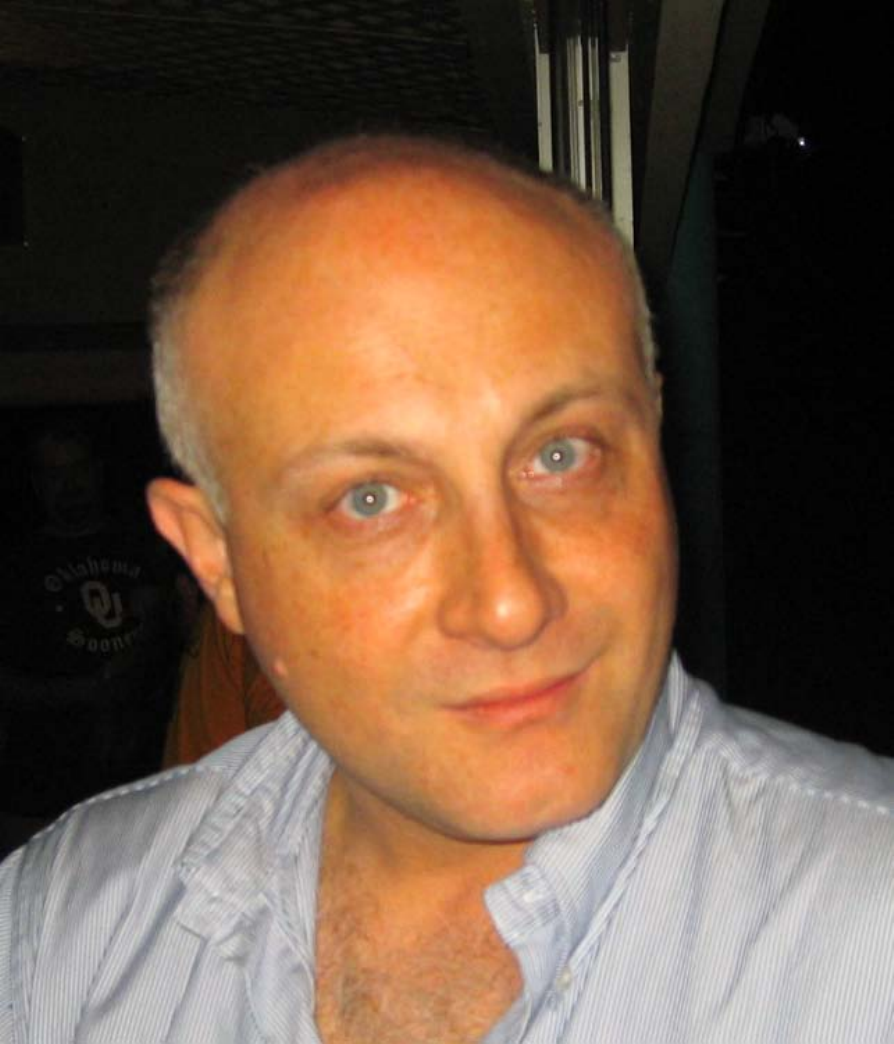}]
{Stefano Galli} (S'95, M'98, SM'05) received his M.S. and Ph.D. degrees in Electrical Engineering from the University of Rome ``La Sapienza'' (Italy) in 1994 and 1998, respectively. Currently, he is the Director of Technology Strategy at ASSIA where he leads the company's overall standardization strategy and contributes to the company's efforts in the area of wired/wireless access and home area networking. Prior to this position, he was in Panasonic Corporation from 2006 to 2010 as Lead Scientist in the Strategic R\&D Planning Office and then as the Director of Energy Solutions R\&D. From 1998 to 2006, he was a Senior Scientist in Bellcore (now Telcordia Technologies).\\
Dr. Galli is serving as elected Member-at-Large of the IEEE Communications Society (ComSoc) Board of Governors and is involved in a variety of capacities in Power Line Communications (PLC) and Smart Grid activities. He currently serves as: Chair of the ITU-T G.hnem project (Narrowband PLC for Smart Grid), Chair of the PAP 15 Coexistence subgroup instituted by the US National Institute of Standards and Technology (NIST), Chair of the IEEE ComSoc Ad-Hoc Committee on Smart Grid Communications, Member of the Energy and Policy Committee of IEEE-USA, and Editor for the IEEE Transactions on Smart Grid and the IEEE Transactions on Communications (Wireline Systems and Smart Grid Communications). He is also the founder and first Chair of the IEEE ComSoc Technical Committee on Power Line Communications (2004-2010), and the past Co-Chair of the ``Communications Technology'' Task Force of the IEEE P2030 Smart Grid Interoperability Standard (2009-2010).\\
Dr. Galli has worked on a variety of wireless and wired communications technologies, is an IEEE Senior Member, holds fifteen issued and pending patents, has published over 90 peer-reviewed papers, has co-authored two book chapters, and has made numerous standards contributions to IEEE and ITU-T.
He has received the 2011 IEEE Technical Committee on Power Line Communications Outstanding Service Award and the 2010 IEEE ISPLC Best Paper Award.
\end{IEEEbiography}

\begin{IEEEbiography} [\addphoto{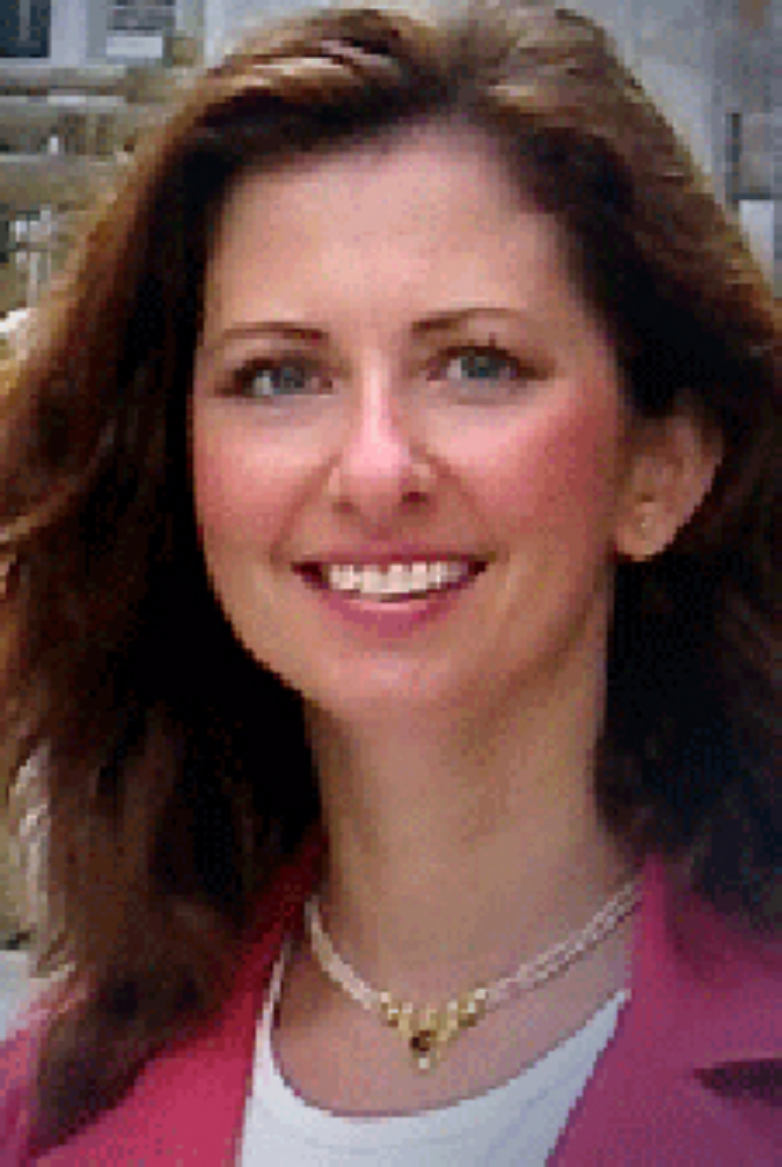}]
{Anna Scaglione}(SM'08, F'11) received her M.S. and Ph.D. degrees in Electrical Engineering from the University of Rome ``La Sapienza", Italy, in 1995 and 1999, respectively. She is currently Professor in Electrical Engineering at the University of California at Davis, CA.  Prior to this she was postdoctoral researcher at the University of Minnesota in 1999-2000, Assistant Professor at the University of New Mexico in 2000-2001, Assistant and Associate Professor at Cornell University in 2001-2006 and 2006-2008, respectively. She received the 2000 IEEE Signal Processing Transactions Best Paper Award, the NSF Career Award in 2002, the Fred Ellersick Award for the best unclassified paper in MILCOM 2005, and the 2005 Best paper for Young Authors of the Taiwan IEEE Comsoc/Information Theory section. Her research is in the broad area of signal processing for communication systems. Her current research focuses on cooperative communication systems and decentralized processing for sensor networks.
\end{IEEEbiography}

\end{document}